%% file: dis98.tex
\documentstyle[sprocl,epsfig,amss,wrapfig]{article}

\input{abbreviations}

\begin{document}
%\pagenumbering{empty}
%
\title{QCD Tests in High Energy Collisions\footnote{
 Summary of Working Group III 'Hadronic Final State',
 6th International Workshop on Deep Inelastic Scattering and QCD,
 Brussels (Belgium), 04-08 April 1998.}
}
\author{\vspace{-0.5cm}
I.~Bertram$^1$, T.~Carli$^2$, G.~Gustafson$^3$, J.~Hartmann$^4$}
\address{
1) Northwestern University, USA \\
2) Max-Planck Institut M\"unchen, Germany \\
3) Lund University, Sweden \\
4) McGill Institute, Montr\'eal, Canada}
\maketitle\abstracts{\vspace{-0.1cm}
 Recent measurements and theoretical developments on 
 the hadronic final state in deep-inelastic scattering,
 \pp and \ee collisions are presented.
 \vspace{-0.8cm}}
%
%\vspace{-0.4cm}
\section{Introduction}
\label{intro}
\input introduction

\mbox{}
\vspace{-1.2cm}
\section{Photons in High Energy Collisions}
\label{photon}
\input photon.tex

\mbox{}
\vspace{-0.8cm}
\section{Heavy Quarks in (real) Photon-Proton Collisions}
%\label{heavyquarks}
\input heavyquarks

%
\section{Jet Shapes}
\label{jetshapes}
\input jetshapes
%
\mbox{}
\vspace{-0.8cm}
\section{Jet Production in Real Photon-Proton Collisions}
\label{multijets}
\input multijets

\mbox{}
\vspace{-1.1cm}
\section{Jet Production in DIS}
\label{dijets}

\input jets

\section{DGLAP / BFKL Parton Evolution Dynamics} 
\label{bfkl}
\input bfkl

\vspace{-0.1cm}
\section{\boldmath High $E_T$ Jet Production in \pp Collisions} 
\label{tevjet}
\input tevjet
%
\vspace{-0.5cm}
\section{\boldmath $W^\pm$ and $Z^0$ Production in \pp Collisions} 
\label{bosjet}
\input bosjet

\mbox{}
\vspace{-0.8cm}
\section*{Instanton Induced Events}
\label{instanton}
\input instanton

\mbox{}
\vspace{-0.8cm}
\section*{Event shapes}
\label{eventshapes}
\input evshapes

\vspace{-0.2cm}
\section*{Fragmentation Functions and Colour Recombination}
\label{fragfunc}
\input fragfun

\vspace{-1.cm}
\section*{Conclusions}
\label{conclusions}
The high statistics available in DIS experiments now allows to
assess details of the parton fragmentation and of QCD effects.
An impressive amount of data is successfully described by
NLO calculations. In particular, in jet production in DIS
significant progress has been recently made.
The transition from non-perturbative to the perturbative regime, 
and the parton dynamics at low $x$ have attracted 
strong interest from experimentalists and from theorists. We are just 
at the beginning of a global understanding for the various interesting
phenomena observed here. 
\section*{Acknowledgments}
We would like to thank all speakers of our session for
an inspiring working group session and 
P. Marage and R. Roosen for the excellent organisation
of the workshop. 
%
%%%%%%%
\section*{References}
\vspace{-0.2cm}
{\small

\input dis98bib
%\bibliography{dis98}
}
\end{document}

%% file: abbreviations.tex
\newcommand{\av}[1]{\mbox{$ \langle #1 \rangle $}}

\newcommand{\xbj}{\mbox{$x_{Bj}~$}}
\newcommand{\xbjx}{\mbox{$x_{Bj}$}}

\newcommand{\xgamma}{\mbox{$x_\gamma^{\rm OBS}~$}}  % x_gamma
\newcommand{\xgammax}{\mbox{$x_\gamma^{\rm OBS}$}}  % x_gamma

\newcommand{\Qsq}{\mbox{$Q^2~$}}
\newcommand{\Qsqx}{\mbox{$Q^2$}}
\newcommand{\et}{\mbox{$E_T~$}}
\newcommand{\etx}{\mbox{$E_T$}}
\newcommand{\etsq}{\mbox{$E_T^2~$}}
\newcommand{\etsqx}{\mbox{$E_T^2$}}

\newcommand{\etonex}{\mbox{$E_{T,{1}}$}}
\newcommand{\ettwox}{\mbox{$E_{T,{2}}$}}
\newcommand{\GeV}{\mbox{\rm ~GeV~}}
\newcommand{\GeVx}{\rm GeV}
\newcommand{\TeV}{\mbox{\rm ~TeV~}}

\newcommand{\GeVsq}{\mbox{${\rm ~GeV}^2~$}}
\newcommand{\GeVsqx}{\mbox{${\rm ~GeV}^2$}}

\newcommand{\nbx}{\mbox{${\rm ~nb}$}}

\newcommand{\pbx}{\mbox{${\rm ~pb}$}}
\newcommand{\pbinv}{\mbox{${\rm ~pb^{-1}}~$}}

\newcommand{\ep}{\mbox{$e^{\pm}p~$}}

\newcommand{\ee}{\mbox{$e^+e^-$}}
\newcommand{\pp}{\mbox{$p\bar{p}~$}}

\newcommand{\costheta}{\mbox{$1-|\cos{\vartheta^*}|$}}
\newcommand{\oalssq}{${\cal O}$$(\alpha_s^2)~$}

%% file: introduction.tex
The hadronic final state in high energy collisions provides
a testing ground for the strong interaction and its underlying
theory QCD. 
The large center of mass energy ($\sqrt{s} \approx 300$\GeVx)
available at the \ep collider  HERA 
allows the production of hadrons in
deep-inelastic scattering (DIS) to be explored in a
new kinematic regime. A wide range of the squared momentum
transfer \Qsq from  $\Qsqx \approx 0 - 10^5$ \GeVsq 
can be accessed. At the same time
very low values of the Bjorken scaling variable
\xbj down to $10^{-5}$ can be reached.

The large data samples %accumulated over the last $5$ years
%corresponding in total to about $40$ \pbinv for each experiment 
do not only allow to study hard processes, but also
focus the interest of experimentalists and theorists on details of 
the complex parton dynamics and the hadronisation of partons 
to observable hadrons. 
Recently, much effort has been put into the calculation of
next-to-leading order (NLO) effects, on a consistent incorporation
of the parton evolution into flexible Monte Carlo programs
and approaches to study power suppressed hadronisation
corrections for specific variables.
%However, the Monte Carlo models, which experimentalists need to correct
%their data for detector effects and for hadronisation
%, are usually leading order only with
%a simulation of higher order effects 
%via a parton shower approach by example.
%The understanding of these simulations is essential for the interpretation
%of the results and many recent studies focus on this importand topics.
%The understanding of the transition region of non-perturbative to 
%perturbative QCD (pQCD). 
%has emerged as one of the most interesting topics in this context.

Results from the \pp collider TEVATRON operating at a
center of mass energy $\sqrt{s}=630 \GeV$ and $1.8 \TeV$ 
and of the $e^+e^-$  colliders LEP I 
$\sqrt{s}=91 \GeV$ and LEP II $130 < \sqrt{s} \lesssim 190$ \GeV
overlap in many kinematic ranges and provide complementary
information. While at the TEVATRON one has access to very high
\et phenomena allowing to probe very short distances,
LEP provides a clean environment to examine QCD effects.
The different center of mass energies from $3$ to $190$\GeV
available in \ee experiments, allow the evolution of QCD  
phenomena % like the running of $\alpha_s$
%or the scaling violation in the fragmentation function
to be investigated~\cite{bethke}.
The unique advantage of HERA is that 
one can probe QCD at different scales in one experiment.
Varying experimental conditions can be realized
only by controlling the scattered electron.
%in one experiment varying experiment conditions, e.g. different hard scales,
%only controlled by the scattered electron can be realised.

Detailed comparison of  hadron production in different reactions 
can be performed and universality of perturbative QCD (pQCD) can be tested.

%% file: photon.tex
In the quark parton model a highly virtual photon 
interacts with a parton freely moving in the proton.
This is a good approximation when small distances are probed, i.e.
in the limit where \Qsq is large.
In this regime,
the photon behaves like a point-like object, i.e. it 
directly couples to quarks to produce the hard scattering.
At low virtualities the photon dominantly 
fluctuates into vector mesons. % with the quantum numbers 
%of the photon. % like $\rho, \omega, \phi$.
%Such interactions can be described within the VDM model~\cite{vdm}.
%When higher transverse energy is available %in the final state,
%photons act as a source of strongly interacting partons 
%e.g. by fluctuating into $q \bar{q}$
%states without forming bound states.
The photon may also fluctuate into a
$q \bar{q}$ state with higher transverse energy without forming a bound
state. In this case the photon acts as a source of strongly interacting
partons. 
Such a process can be calculated within pQCD using the concept 
of a photon structure function.
%%%%%%%%%%%%%%%%%%%%%%%%%%%%%%%%%%%%%%%%%%%%%%%%
\begin{figure}[t]
%\rule{5cm}{0.2mm}\hfill\rule{5cm}{0.2mm}
\vspace{-0.2cm}
\mbox{\hspace{-0.5cm}
\begin{tabular}{cc}
\epsfig{figure=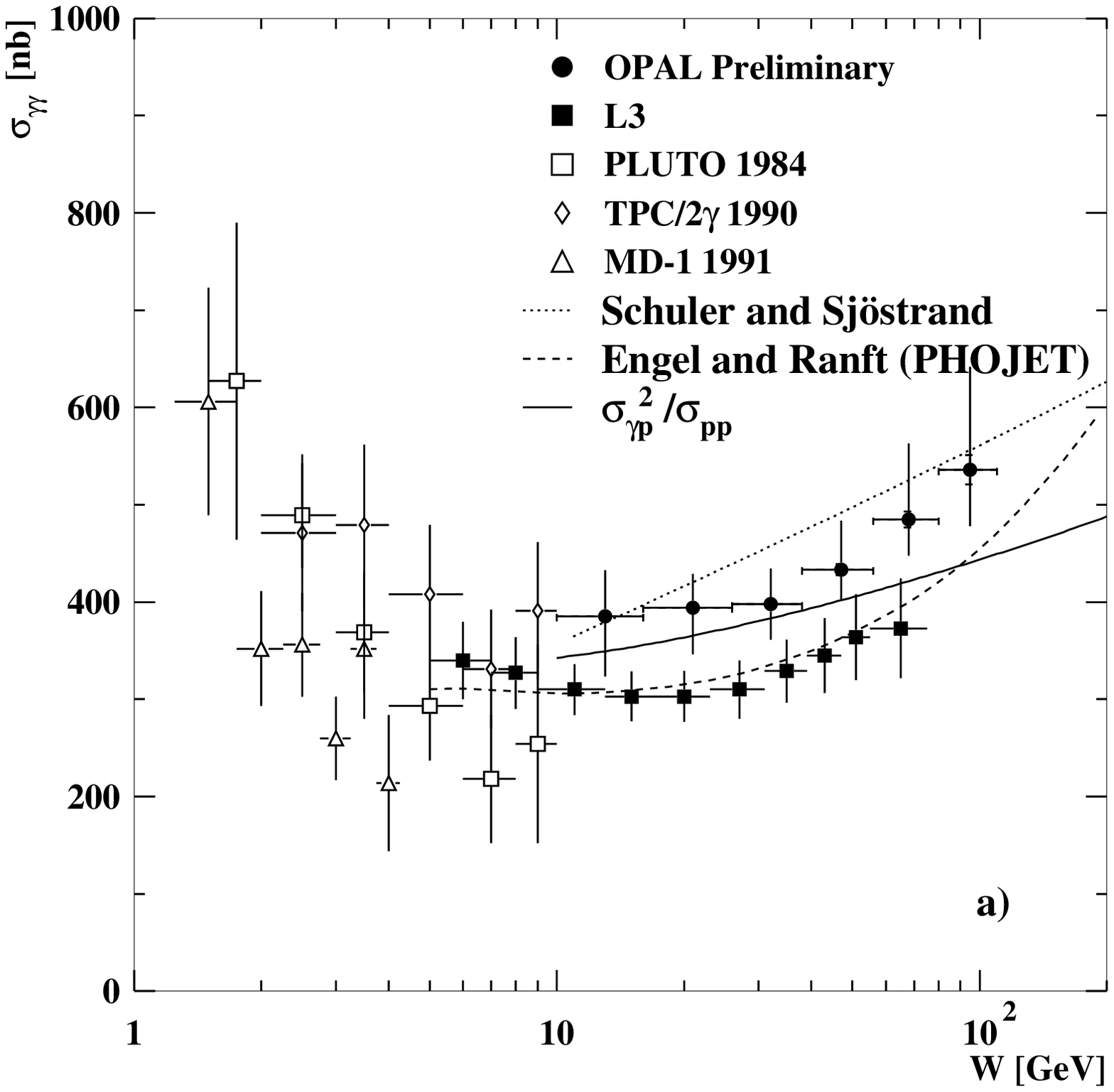,width=6.cm}
\mbox{\hspace{-0.5cm}
\epsfig{figure=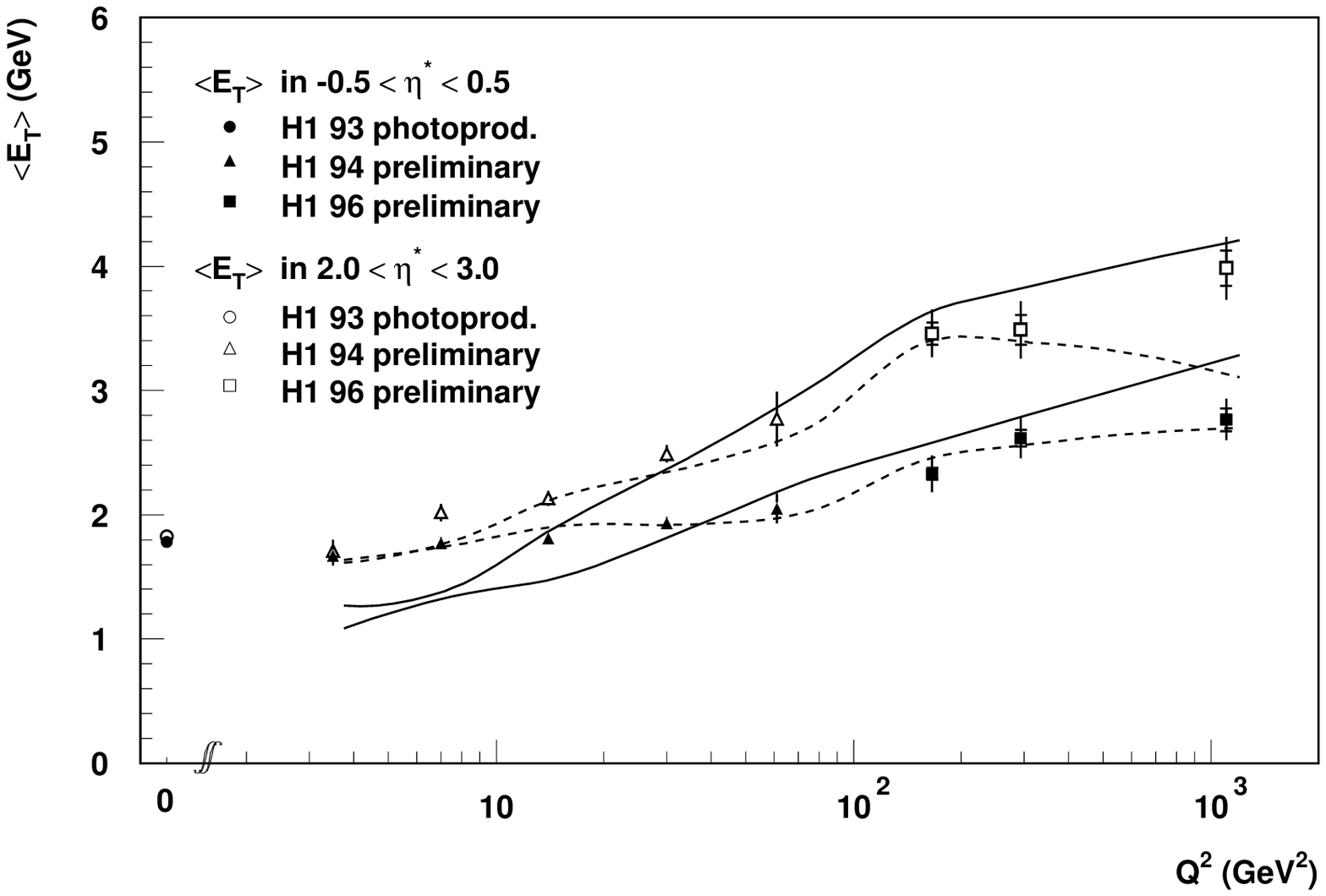,
bbllx=24,bblly=430,bburx=532,bbury=749,clip=,width=6.5cm,height=5.5cm}
}
\end{tabular}
}
\vspace{-0.2cm}
\caption{
a) Total cross-section $\sigma_{\gamma,\gamma} \to hadrons$ %~\cite{Buergin} 
   as a function of the invariant mass of the hadronic final state $W$.
b) $\av{\etx}$ in two rapidity regions in the hCMS as a function of $Q^2$.
%   in the central rapidity 
%   region $-0.5 < \eta < 0.5$ and in the photon fragmentation region 
%   $2 < \eta < 3$ in the hCMS as a function of $Q^2$.
   Overlayed are the Monte Carlo predictions from LEPTO (solid) 
   and CDM (dashed).      
}
\label{fig:photon}
\end{figure}
%\clearpage
%%%%%%%%%%%%%%%%%%%%%%%%%%%%%%%%%%%%%%%%%%%%%

Photon-photon interactions at high center of mass energies 
can be studied at LEP I and LEP II using
doubly anti-tagged $e^+e^-$ scattering events.
They provide the main source of hadron production above the $Z^0$ mass.
The total cross section $\sigma_{\gamma\gamma}\rightarrow hadrons$ 
is measured by L3 and OPAL~\cite{burgin} (see Fig.~\ref{fig:photon}a). 
The LEP II data together with earlier results~\cite{lowW} at small
energies show the characteristic slow 
rise expected from the hadronic nature of the photon for
increasing invariant masses $W$ of the hadronic final state
in the event. Such a rise is  consistent with the universal
behaviour of the total cross-section~\cite{donnachie}
in $\gamma p$ scattering at HERA~\cite{sigtothera}
and in \pp collisions at TEVATRON~\cite{tevpp}.
It is well described by QCD models~\cite{sigggmodels}.
The two recent measurements show a similar
$W$ dependence, but disagree in the absolute 
values (see Fig.~\ref{fig:photon}a).

Events where the perturbatively calculable components become
more important can be selected by tagging heavy quarks
in the final state. The cross-section $e^+ e^- \to e^+ e^- c \bar{c}$
measured~\cite{kienzle} as a function of $\sqrt{s}$ is well
described by a full NLO calculation~\cite{drees}. A
substantial resolved contribution is required.

In DIS, the hadronic nature of the photon can be illustrated
in the reference frame where the
proton is at rest and the photon is fast.
In this frame the photon fluctuates - before 
interacting with the proton - into a complex object
with a typical size of $c \tau_\gamma=200$ {\rm fm}
at $\xbjx \approx 10^{-3}$. %, which is large to the proton
%radius of $1$ {rm fm}.
For suitable observables one can find characteristic properties
of hadron-hadron collisions in DIS at low-\xbjx.
When comparing the \av{\etx} produced in the central 
pseudo-rapidity bin 
$ |\eta| < 0.5$ of the hadronic center of mass system (hCMS)
to hadron-hadron collisions, it scales with the available
center of mass energy ($\sqrt{s}$ or $W$)
independent of the nature of the incoming
particle~\cite{h1:mult}.
In an early publication~\cite{h1:eflcomp}, the H1 collaboration has shown
that at fixed $W=180$ \GeV 
the \av{\etx} in $ |\eta| < 0.5$
is independent of the photon virtuality.
At this conference, new results in an extended
kinematic range and with a better understanding of systematics
effects have been presented~\cite{kruecker}.
The $\av{\etx}$ in the central bin is constant for 
$0 \lesssim \Qsqx \lesssim 50$\GeVsqx, but then rises 
(see Fig.~\ref{fig:photon}b).
%This is surprising, 
Since the typical short range correlation in 
inelastic hadron-hadron collisions
is about $\Delta\eta \lesssim 2$, the \et induced by the photon virtuality 
should become negligible in this region. 
The unexpected rise might be associated with
a kinematic shift of the highest \et position as a function of \Qsq
(see Fig.~$3$ in ref.~\cite{kruecker}).
%
%(about $\eta=2$ at \av{\Qsqx}=175 \GeVsq and
%$\eta=1$ at \av{\Qsqx}=2200 \GeVsqx.
It would be interesting to see, if
after a redefinition of the fragmentation and central rapidity bin,
the expected constant behaviour of the \av{\etx} %as a function of \Qsq
can be verified in the data.

%% file: heavyquarks.tex
At HERA, heavy quarks are dominantly produced in real
$\gamma$-p collisions. They can be identified via tagging
hadrons forming bound states of heavy quarks alone (hidden)
or heavy and light quarks (open). 
%Hidden charm and beauty
%has been discussed in the diffractive working group.

%The production of heavy quarks naturally offers a hard scale
%through which  reliable pQCD calculations can be expected.
Recent NLO calculation for charm production apply two
schemes. In the 
massive scheme~\cite{nlomassiv} 
the charm mass provides a natural hard scale and charm
quarks only appear in the hard subprocess, this means
that they are not an active flavour in the parton evolution.
In the massless scheme~\cite{nlomassless} charm is treated 
like any other light quark.
Terms proportional to $\alpha_s \ln(p_t^2/m_c^2)$ in the parton evolution
are summed to all orders, but charm mass effects are ignored. 
This approach should give a better
description of data, if $p_t^2/m_c^2$ is large.
To compute the $D^*$ cross-section one needs an appropriate 
fragmentation function, like the Peterson function~\cite{peterson}, 
which is constrained by \ee data. 
Another approach %to calculate heavy quark production 
is to complement a LO matrix element by higher order contributions
based on the BFKL evolution of an unintegrated gluon
distribution~\cite{zotov}.

%%%%%%%%%%%%%%%%%%%%%%%%%%%%%%%%%%%%%%%%%%%%%%%%%%
\begin{figure}[htb]
\vspace*{-0.6cm}
\begin{tabular}{cc}
\mbox{\hspace{-0.5cm}
      \epsfig{figure=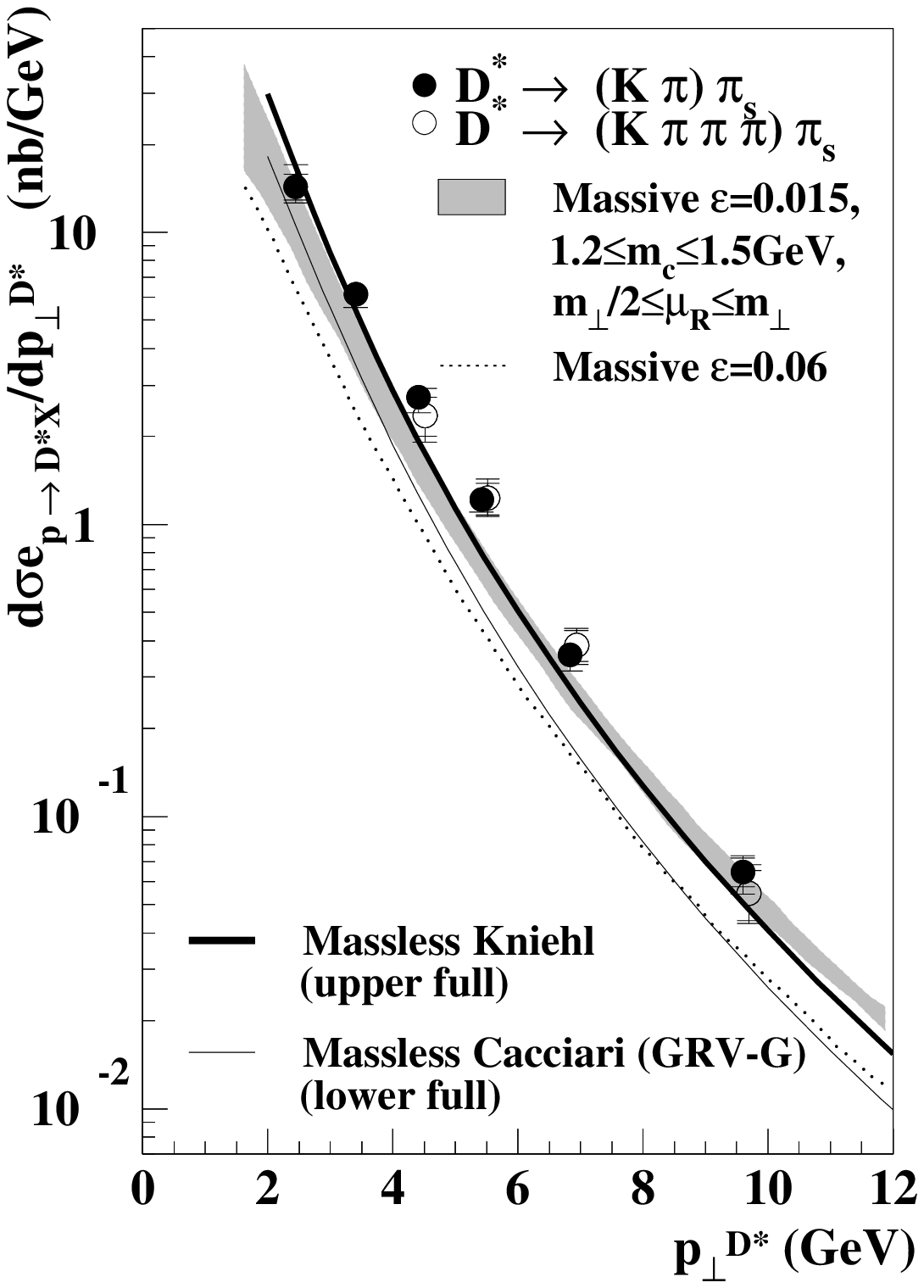,width=6.cm}
     } &
\mbox{\hspace{-0.7cm}
      \epsfig{figure=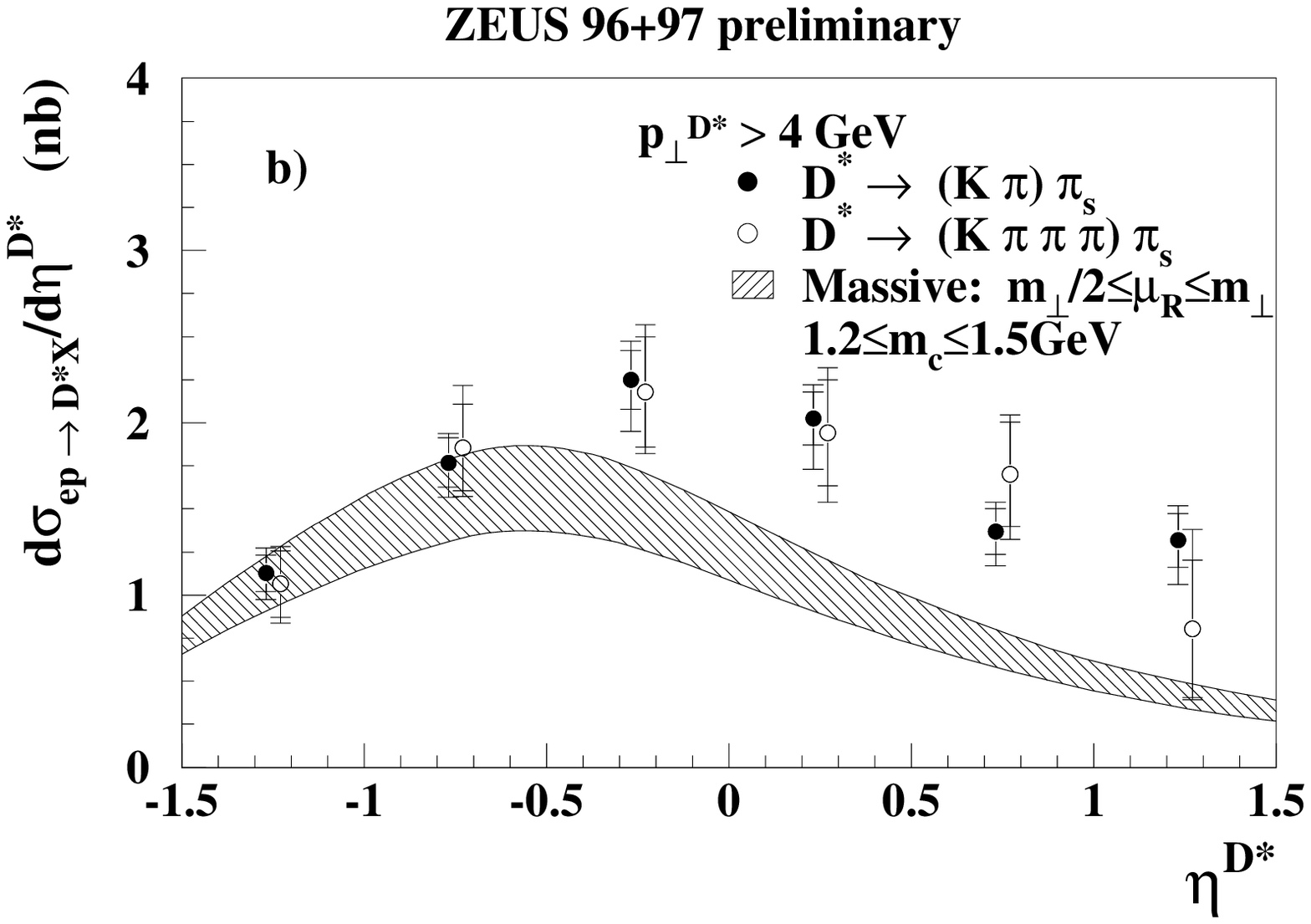,width=7.cm,height=7.cm}
     }
\end{tabular}
\vspace*{-0.9cm}
\caption{$D^*$ cross-section versus
         a) the trans. momentum and 
         b) the pseudorapidity.}
\label{fig:Sutton_1}
\end{figure}
%%%%%%%%%%%%%%%%%%%%%%%%%%%%%%%%%%%%%%%%%%%%%%5

The differential cross-section $d\sigma/dp_t$ for $|\eta|<1.5$
and $d\sigma/d\eta$ for $p_t>4$\GeV measured for two $D^*$ decay 
channels~\cite{sutton} is shown in Fig.~\ref{fig:Sutton_1}. 
None of the NLO approaches is able to described the data.
The discrepancy is most pronounced in the direction towards
the proton remnant. This is, however, the region where the
variation of theory parameters like the
gluon density in the photon, the charm mass, the renormalisation
scale or the fragmentation parameters (e.g. $\varepsilon$) 
have the smallest influence.
To get more information on event kinematics
%the nature of the discrepancy
a subsample containing hard dijets with 
$\etonex > 6$ and $\ettwox > 7$\GeV and $|\eta^{jet}| < 2.4$
is selected. %This allows to get further information on the
%event kinematics. 
A comparison of $d\sigma_{\rm dijet}/d\xgamma$
to a LO QCD model shows a significant contribution from resolved
events (in a LO interpretation) at low $\xgammax < 0.75$.
In contrast to the inclusive dijets analysis, 
%(see section \ref{multijets}),
a LO QCD model is able to describe the absolute cross-section.
A massive NLO calculation fails at low $\xgamma$ to describe the data.
It will be interesting to compare the data also to a massless scheme
which should give a better description at high $p_t$.
 
Open $b$ production has been studied for the first time at HERA
using the semi-leptonic decay channels.
Dijet events accompanied by at least one muon within one jet 
in the final state are selected.
The probability that (misidentified) muons origin from other hadrons
like pions, kaons or protons is extracted from the data and
cross-checked by a detector simulation.
The content of fake muon in the data sample is 
about $24\%$.
The beauty and charm contribution  
is deduced on a statistical basis
exploiting that the $p_t$ of the muon relative to the jet
thrust axis is higher for $b$ events. A relative fraction of
$\approx 52 \%$ for beauty and $24\%$ for charm is found.
The measured visible cross-section for $0.1< y < 0.8$, $p_t^\mu > 2$ \GeV
and $35 < \theta^\mu < 130^o$ is 
$0.93 \pm 0.08 (stat.)^{+0.21}_{-0.12} (syst.) \nbx$. 
This cross-section is about $5$ times larger than 
the LO QCD model AROMA~\cite{aroma} prediction.
NLO calculations~\cite{bnlo} are not yet available for the specific
analysis cuts. However, corrections not higher than a factor of $2$
are expected.
%It is interesting to note that 
The unexpected high $b$ cross-section
might explain part of the $D^*$ 
excess in the forward region.
$b$ events are produced more forward
(see Fig.~$2b$ in ref.~\cite{tsipolitis}) and the $D^*$
appears in a large fraction of the $b$ decay channels.

%$b$ decays with a $D^*$ in the final state contribute
%a large fraction 
%$D^*$ originating from $b$ events are produced more forward
%(see Fig.~$2b$ in ref.~\cite{tsipolitis}).

%% file: jetshapes.tex
%%%%%%%%%%%%%%%%%%%%%%%%%%%%%%%%%%%%%%%%%%%%%%%%
\begin{wrapfigure}[10]{r}{8.5cm}
%\begin{figure}
\vspace{-1.9cm}
\epsfig{figure=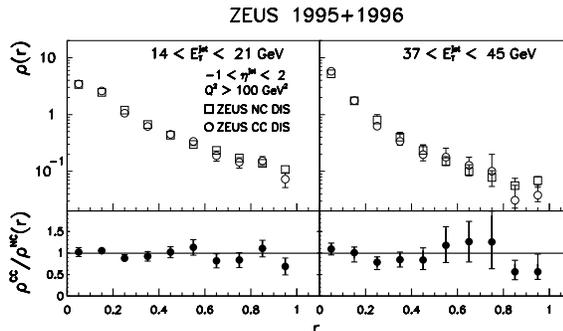,width=8.5cm}
\vspace{-1.6cm}
\caption{Jet shapes in neutral current and charged current events
         in two different \et bins  %~\cite{Martinez}
         in the lab. frame for $\Qsqx > 100$\GeVsqx.
}
\label{fig:Martinez}
\end{wrapfigure}
%\end{figure}
%%%%%%%%%%%%%%%%%%%%%%%%%%%%%%%%%%%%%%%%%%%%%
The internal structure of jets provides
useful information on the transition of a parton to the
complex aggregate of observable hadrons. Since jets
are used as manifestations of hard partons in many studies 
of pQCD, it is important to prove that their detailed properties 
can be well described.

The (differential) jet shape $\Psi(r)$ ($\rho(r)$)
is defined as the average fraction of the \et of the 
jets inside an inner cone of radius $r$
(inside a cone slice with width $\Delta r_0$) 
concentric to the outer jet cone with radius $R = 1$:
\vspace{-0.2cm}
\begin{equation}
\Psi(r) = \frac{1}{N_{\rm jet}} \sum_{\rm jets}
\frac{\etx(0,r)}{\etx(0,R)}
\; \; ,  \; \;
\rho(r)= \frac{1}{N_{\rm jet} }  
\sum_{\rm jets} \frac{\etx(r-\Delta r_0/2,r+\Delta r_0/2)}{\Delta r_0 \; 
\; \; \; \etx(0,R)},
\vspace{-0.2cm}
\end{equation}
where $N_{\rm jet}$ is the number of jets.
By definition $ 0< \Psi < 1$ and
$\Psi(R)=1$. The steepness of the rise (fall) of $\Psi$ ($\rho$) 
describes the collimation of the jet. 

Jet shapes corrected for detector effects
have been measured by ZEUS~\cite{martinez,zeusjetshapes} 
for DIS at $\Qsqx > 100$\GeVsq in the laboratory frame
using jets with $\etx_{,{\rm lab}} > 14$\GeV 
and $-1 < \eta_{\rm lab} < 2$. 
The differential jet shape peaks around $0$ and falls
down by a factor of $40$ towards the edge of the 
jet (see Fig.~\ref{fig:Martinez}).
Jets become narrower as \et increases. No $\eta$ dependence is found.
Jets produced in neutral and charged current interactions
as well as in \ee interactions~\cite{eejetshape} behave 
in a similar way.

The H1 analysis~\cite{wobisch} is performed at 
$10 < \Qsqx <120$\GeVsqx. Jets with $\etx > 5$\GeV
and $ -1 < \eta_{\rm lab} < 2$ are selected 
in a dijet sample in the Breit frame. 
%%%%%%%%%%%%%%%%%%%%%%%%%%%%%%%%%%%%%%%%%%%%%%%%
\begin{wrapfigure}[12]{r}{8.2cm}
\vspace{-0.5cm}
% \mbox{\hspace{1.0cm}
 \epsfig{figure=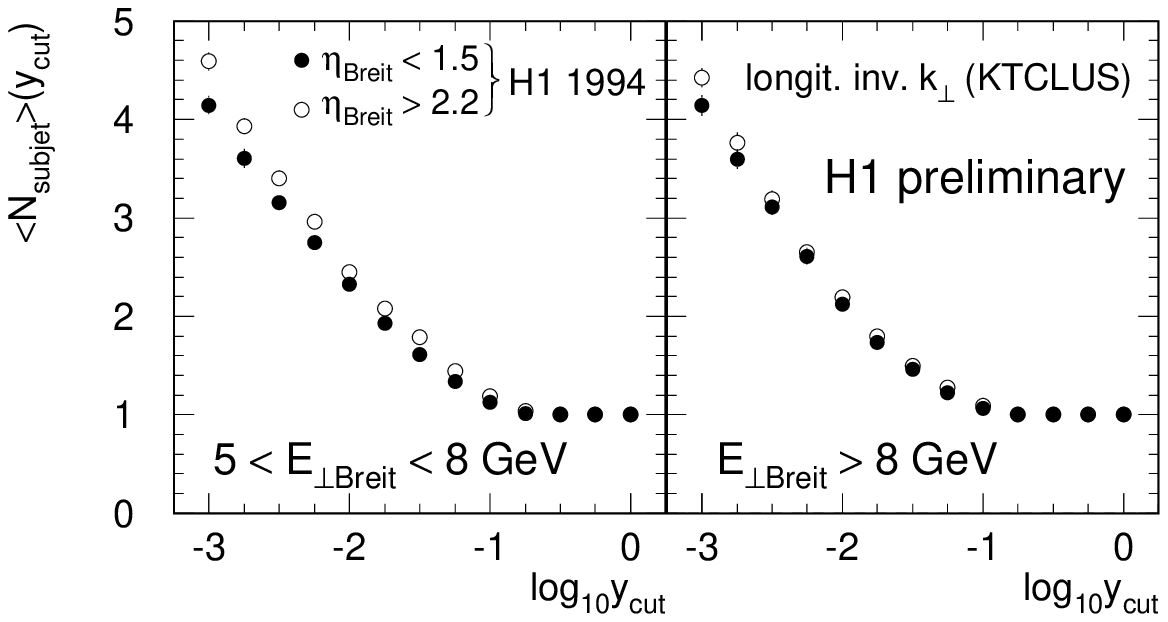,width=9.cm}
% }
\vspace{-1.cm}
\caption{Mean subjet multiplicity %\cite{Wobisch} 
 for dijets at low \Qsq with $\etx > 5$~\GeV
 as a function  $y_{cut}$.
}
\label{fig:subjets}
\end{wrapfigure}
%%%%%%%%%%%%%%%%%%%%%%%%%%%%%%%%%%%%%%%%%%%%%
%Jets get narrower with increasing $E_T$.
In the phase space region chosen by H1, jets get less collimated towards 
the proton remnant direction.
Another interesting observable is the mean number of subjets
defined by a repetition of the
clustering procedure\footnote{Jets are here defined by the
longitudinal invariant $k_t$ algorithm 
for DIS~\cite{invkt}.} within a selected jet.
The clustering is stopped when all particles are above
some cut-off $y_{\rm cut}$. Remaining particles are called subjets.
As is shown in Fig.~\ref{fig:subjets} the number of subjets
decreases for higher \et (for $\eta > 2.2$) and in particular for low \et
more subjets are found towards the target region.

%$\to$ difference quark gluon jets in ee from Bentvelsen

%% file: multijets.tex
 Hard jets allow precise tests of basic QCD predictions.
 The largest event samples of high \et jets at HERA
 are available in (real) photon
 proton collisions. Many subprocesses with different
 angular distributions %in the hCMS contribute 
 contribute to the total dijet cross-section.
 In the resolved channel gluon exchange 
% ($qg \to q g$) 
 dominates and leads to a 
 $(\costheta)^{-2}$ distribution\footnote{
 $\vartheta^*$ is the angle between the jet with the highest 
 \et and the beam direction in the center-of-mass system
 of the hard subprocess.}.
 For events where the photon couples directly,
 a quark is dominantly exchanged. % ($q \gamma \to q g$).
 The angular distribution is then of the form $(\costheta)^{-1}$.
 Events from resolved processes are therefore in LO 
 $\alpha_s$ expected
 to rise more steeply %with increasing \costheta~ 
 towards small $\vartheta^*$
 and the relative contribution of direct to resolved
 processes should increase with increasing jet \et 
 (since ${\cos{\vartheta^*}}= 1 - 4 \, \etsqx/s$).

% Also: at low Pt cross-section diverges -> need hard jets
% to test pert. QCD set eg. alpha_s to ptjet as scale
% 
% $\cos{\theta^*}$ can be calculated from the rapidity difference
% of the jets with the highest \et.
 Resolved and direct processes can be operationally defined
 by a cut on $\xgamma$ defined by %the ratio of $E-P_z$ summed
 %over the two hard jets and all particles.
 $\sum_{jet} (E - P_z) / \sum_{had.} (E - P_z)$.
 In LO, \xgamma can be interpreted as
 the fraction of the longitudinal
 momentum carried by the interacting parton in the photon.
 It should be $1$ for direct and smaller for resolved processes.
 In NLO the direct and resolved components of the total 
 cross-section are strongly correlated.
 The operational definition ($\xgammax < 0.75$) performed in the
 experimental analysis to enhance the resolved component
 leads to a sizeable mixture between NLO point-like and resolved components.
%%%%%%%%%%%%%%%%%%%%%%%%%%
\begin{wrapfigure}[17]{r}{7.5cm}
\vspace{-1.1cm}
\epsfig{figure=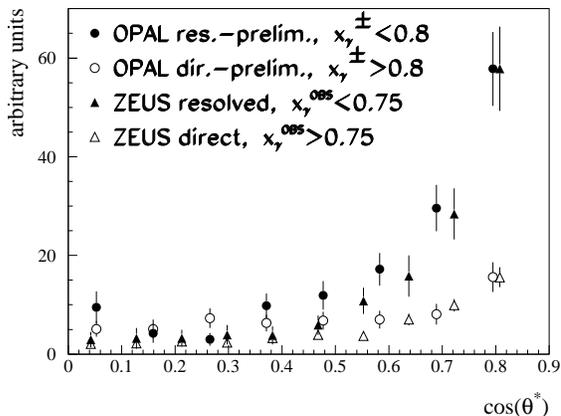,width=8.cm}
\vspace{-1.1cm}
\caption{Distribution of the polar angle of the highest \et jet
         with respect to the beam in the rest frame of the $2$ jet system.
} 
\label{fig:cos_theta}
\end{wrapfigure}
%%%%%%%%%%%%%%%%%%%%%%%%%%%%%
 
 In Fig.~\ref{fig:cos_theta} the angular distribution for
 dijet events produced in \ep~\cite{ZEUSdijets} 
 and \ee collisions~\cite{burgin} 
 are shown\footnote{In the case of \ee: 
 $x_{\gamma}^\pm = \sum_{\rm jet} (E \pm P_z) / \sum_{\rm had.} 
(E \pm P_z) <0.8$
 is used.}. 
 The qualitative behaviour known from LO
 arguments is seen in the data.
 It has been, however, pointed out~\cite{frixione}
 that in a NLO calculation the characteristic \costheta~
 distribution is not a result of the
 different propagator in resolved and direct processes (both
 distributions are flat), but are generated by the kinematical cuts
 $M_{jj}>23 \GeV$ and $|{\eta}| <0.5$ applied in the 
 \ep analysis\footnote{Similar cuts have been applied in the
 \ee analysis: $M_{jj}>12 \GeV$ and $|{\eta}| < 2$.}. 
 The steeper rise for low $\xgamma$
 is induced by a softer $M_{jj}$ distribution.
 It is therefore not clear, if an operational definition is
 useful to extract information on the underlying dynamics.

\begin{figure}[t]
\mbox{\hspace{-0.5cm}
 \begin{tabular}{cc}
\epsfig{figure=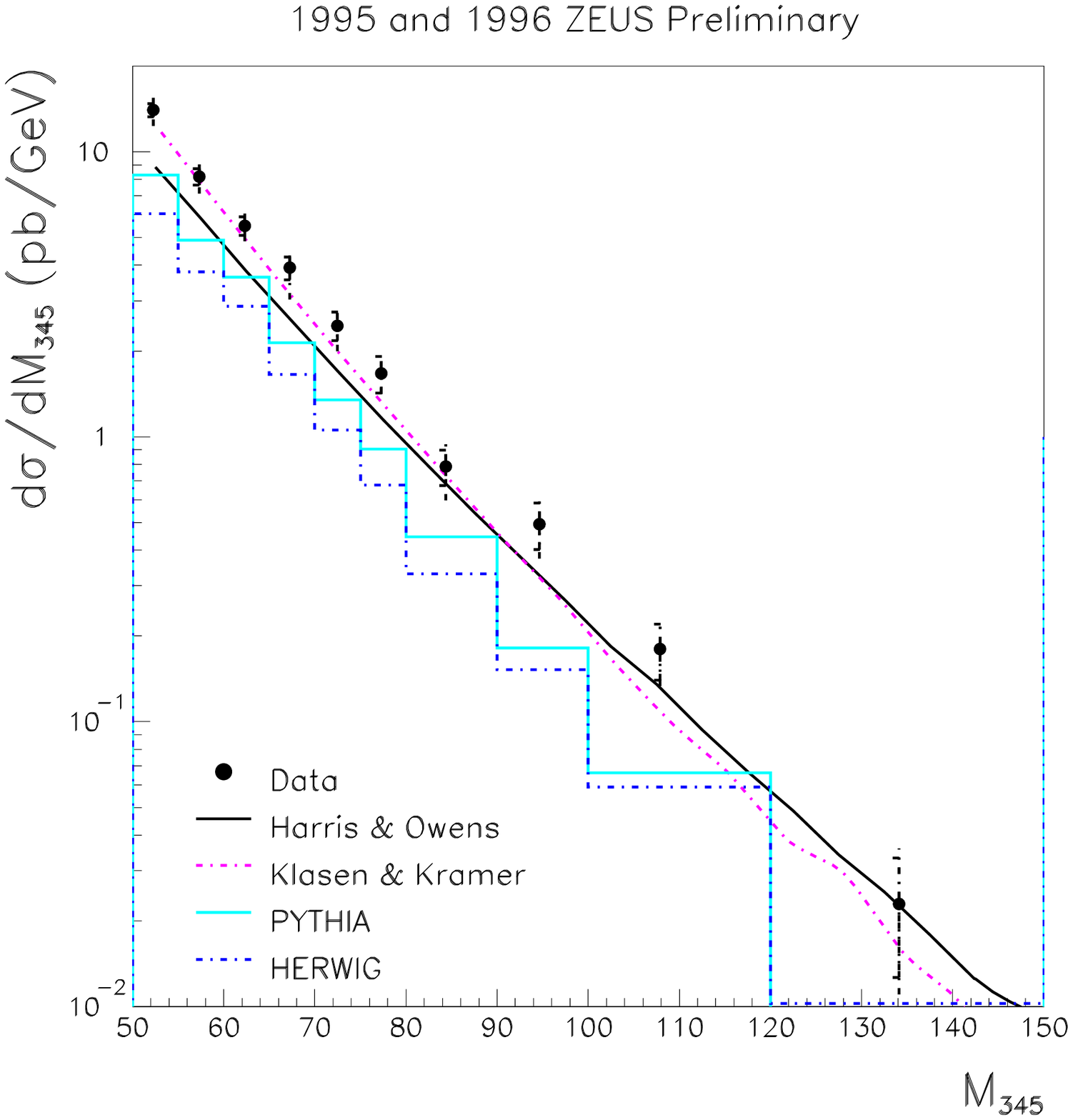,width=5.cm}
\mbox{\vspace{0.6cm} 
\epsfig{figure=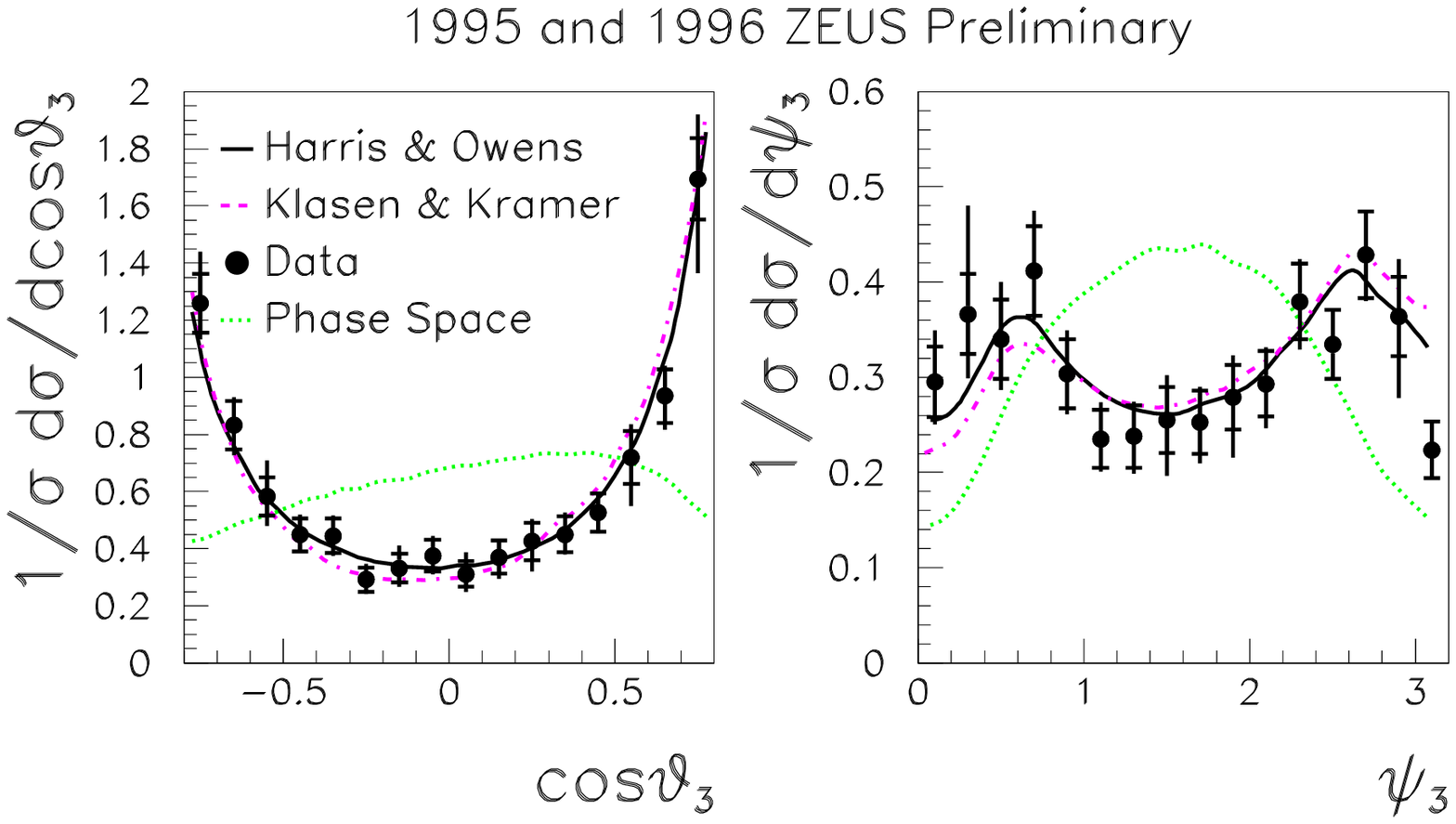,width=7.cm,height=5.8cm}
}
\end{tabular}
}
\vspace{-0.3cm}
\caption{
a) Invariant mass of the 3-jet system. %~\cite{Sinclair}. 
b) Angular distributions for jets with respect to certain planes
of the event, see text for details.}
\label{fig:Sinclair}
\end{figure}

 Properties of events with three jets can be compared to 
 \oalssq QCD calculations~\cite{klasen,harris} 
 which represent the LO for such processes.
 A basic quantity is the invariant mass of the three jet system $M_{3j}$.
 The  $M_{3j}$ distribution~\cite{sinclair} shown in 
 Fig.~\ref{fig:Sinclair} is reasonably well described by
 \oalssq calculations.
 The LO QCD models 
% complemented with parton showers to account for leading $\log{\Qsqx}$ 
 like PYTHIA~\cite{PYTHIA} or HERWIG~\cite{HERWIG}
 fail in absolute, but are able to reproduce the shape.
%%%%%%%%%%%%%%%%%%%%%%%%%%%%%%%%%%%%%%%%%%
%%%%%%%%%%%%%%%%%%%%%%%%%%%%%%%%%%%%%%%%%
%
 In three jet events, in addition to the polar angle 
 $\cos{\vartheta_3}$ of the
 jet with the highest \et and the beam direction, $\Psi$,
 the angle between the plane containing the three jets and the plane
 containing the highest \et jet and the beam direction, can be defined.
 The measured $\cos{\vartheta_3}$ is different from what is
 expected from pure phase space arguments (see Fig.~\ref{fig:Sinclair}b).
 \oalssq QCD calculations
 and the LO QCD models give a good description
 of the data. Due to the selection cuts
 the phase space near $\Psi \approx 0$ and $ \Psi \approx \pi$
 is artificially suppressed. Nevertheless, a
 strong tendency for the three jet plane to lie near the plane containing
 the beam and the highest \et jet can be seen in the data.
 Such a characteristic behaviour is expected from the collinear singularities
 of QCD radiation.

%% file: jets.tex
The production rate of dijet events in DIS is directly
sensitive to the QCD free parameters like the strong
coupling constant $\alpha_s$ or the parton density functions in the proton.
The possibility to perform NLO precision tests has stimulated a lot
of activity since the beginning of 
HERA~\cite{alphash1,alphaszeus,flamm}.
Early analyses were restricted to semi-analytical
NLO calculations~\cite{projet,disjet} which were only available 
for the JADE algorithm~\cite{jadejet}.
These calculations did, however, not use exactly the same
jet definitions as in the experimental analysis~\cite{rosenbauer}. 
Moreover, approximations were made
which turned out not to be valid over the full phase
space region~\cite{mirkes1}. Meanwhile the flexible 
NLO Monte Carlo programs MEPJET~\cite{mirkes1}, 
DISENT~\cite{disent}, DISASTER++~\cite{disaster} and JETVIP~\cite{jetvip} 
became available which allow arbitrary jet definition schemes and
experimental cuts. The H1 collaboration has now
updated their results\footnote{The ZEUS results do not change
when using the new NLO calculations~\cite{trefzger}.} 
on the $\alpha_s$ determination 
from integrated and differential jet rates~\cite{tobien} 
using the JADE algorithm. 
 
At the DIS~$97$ conference, agreement of NLO calculations
with the data could only be achieved in very restricted 
phase space regions~\cite{jetdis96}.
Especially in the region of low \Qsqx, both NLO calculations
and LO QCD models failed to describe
the data. This had been interpreted as a sign of a contribution
of (LO) resolved photons in DIS~\cite{jungresolved}, but it was not clear
why the shape of all distributions associated to the hard subprocess
were well described by NLO calculations. 
%%%%%%%%%%%%%%%%%%%%%%%%%%%%%%%%%%%%%%%%%%%%%%%%
\begin{wrapfigure}[27]{l}{7.cm}
%\begin{figure}[t]
%\begin{tabular}{cc}
%\mbox{\hspace{-0.4cm}
\vspace{0.3cm}
\epsfig{figure=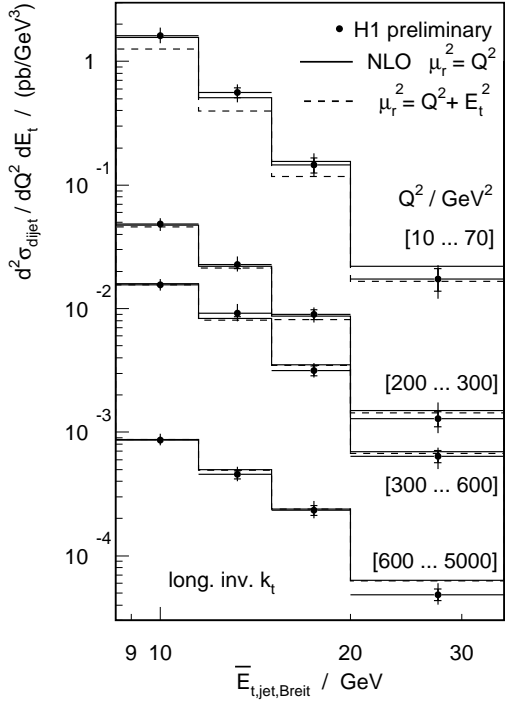,width=7.cm}
%\epsfig{figure=delta_q2_1.eps,width=6.cm,height=7.cm}
%\mbox{\vspace{-0.5cm}
%\epsfig{figure=r2delta.eps,width=6.cm,height=9.cm}
%}}
%\end{tabular}
\vspace{-0.9cm}
\caption{
 Double differential dijet cross-section 
   in the Breit system
   NLO calculations with different choices of the
   renormalisation scales are overlayed.
%b and c) Dijet rate as a function of the difference of the jet 
%   \et thresholds ($\Delta$) for two different \Qsq bins.
}
\label{fig:wobisch}
%\end{figure}
\end{wrapfigure}
%%%%%%%%%%%%%%%%%%%%%%%%%%%%%%%%%%%%%%%%%%%%%

Meanwhile, it has been understood that this discrepancy
only occurs when both jets are required to be above
the same \et threshold. 
%The NLO prediction\footnote{
%\Qsq is used as renormalisation scale.} 
%for the dijet rate as function
%of $\Delta$, the required \et difference of the jet thresholds,
%is shown in Fig.~\ref{fig:wobisch}b and c together with H1 data points
%for two \Qsq bins.
As $\Delta$, the required \et difference of the jet thresholds,
approaches $0$,
a fixed order calculation gets infra-red sensitive~\cite{frixione}.
% , since 
%the emission of a soft gluon can reduce the transverse energy
%of one of the jets such that it falls below the threshold.
In events where the \et of the jets are approximately equal, 
no phase space to emit
a third real parton is available. This leads to an incomplete
cancelation between real and virtual corrections
at the threshold and makes a fixed order calculation 
unpredictive~\cite{frixione,poetter}. 
A resummation of higher orders
is necessary at this phase space point. First theoretical attempts
for resummation of dijet production in hadron hadron collisions
have been presented at this conference~\cite{kidonakis}.
The updated dijet rates~\cite{wobisch} for $5 < \Qsqx < 100$~\GeVsqx, 
where $\Delta > 2$ or
$\etonex + \ettwox > 13 $~\GeV is required,
are well described by NLO calculations.

%Based on the experienced gained on the dijet rates, 
The H1 collaboration has presented 
a comprehensive measurement~\cite{wobisch} of the double differential dijet
cross-section corrected for detector effects
from $10 < \Qsqx < 5000$~\GeVsq and 
$8.5 < \av{\etx} \lesssim 35$~\GeVx.
Jets defined in the Breit frame with several 
jet algorithms~\cite{invkt,ktalg,cambridge} 
are required to have $\etx > 5$~\GeV and $\etonex + \ettwox > 17$~\GeVx. 
%Jets are found using the
%long. inv. $k_t$ algorithm~\cite{XXXX} requiring $et>5$ \GeV and
%$\etx_1+\et_2> 17$ \GeVx.
The interplay between the two hard scales \Qsq and \et can be
seen in the data (see Fig.~\ref{fig:wobisch}). The \av{\etx} spectra of the
jets get harder for increasing \Qsqx.
A NLO calculation~\cite{disent} provides a
good description over the whole kinematic range\footnote{
Hadronisation corrections are below $7\%$ for 
$\Qsqx > 200$ \GeVsq and below $15\%$ for lower \Qsqx.}.
These are the first encouraging steps
towards a fundamental understanding of 
dijet production in DIS at HERA.
At low \Qsq NLO corrections are large. 
%In the region towards the
%proton remnant they reach a factor $5$. 
Moreover, they depend on the choice of
the renormalisation scale.
A good description of the data can only be achieved using \Qsq
as scale. The conceptually preferred scale
$\Qsqx + \etsqx$ providing a natural interpolation between
the Bjorken and the photoproduction limit
falls below the data.
It is an open question whether the success of the \Qsq scale
is accidental and whether the failure of the $\Qsqx + \etsq$ scale
is a sign that higher orders can give significant contributions
at low \Qsqx.

The question to which extent contributions from resolved
photons are needed to describe the data can be addressed 
using the NLO program JETVIP~\cite{jetvip} incorporating the 
NLO Matrix elements together with a virtual photon structure function.
Higher order emissions in the photon
direction beyond the one which is already included in 
the NLO matrix elements (mainly the diagram where the photon splits
into a $q\bar{q}$ pair) are only needed in the region of 
low \Qsqx.

%% file: bfkl.tex
%%%%%%%%%%%%%%%%%%%%%%%%%%%%%%%%%%%%%%%%%%
\begin{wrapfigure}[14]{r}{6.0cm}
\vspace{-1.2cm}
\epsfig{figure=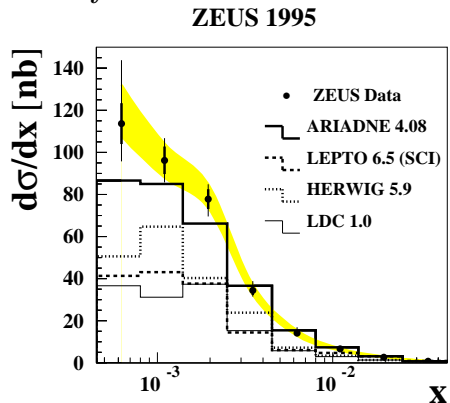,width=7.0cm} 
\vspace{-2.2cm}
\caption{Forward jet cross sections %~\cite{Riveline} 
         as a function of $x_{Bj}$.}
\label{fig:Riveline_1}
\end{wrapfigure}
%%%%%%%%%%%%%%%%%%%%%%%%%%%%%%%%%%%%%%%%%%%%%

Several talks concentrated on the parton evolution dynamics at low 
$x_{Bj}$~\cite{wengler,Riveline,Salam,Lonnblad,Stirling}.
%The two existing schemes, DGLAP~\cite{DGLAP} and BFKL~\cite{BFKL}, 
%should start to deviate from each other at low \xbjx, due to the 
%$\ln{1/x_{Bj}}$ terms, which are omitted in the DGLAP equations.
The two existing schemes, 
DGLAP~\cite{DGLAP} and BFKL~\cite{BFKL}, describe the evolution towards
large values of \Qsq and $1/\xbj$ respectively. 
Thus for low \xbj and moderate
\Qsq the $k_t$-ordered DGLAP evolution should be inappropriate.
A solution of the parton evolution equation by CCFM~\cite{CCFM} 
approximates the BFKL equation in the low \xbj limit and the DGLAP
prediction in the high \xbj limit. 

%While theorists perform calculation for partons, 
%experimentalists prefer to correct their data for detector effects only, 
%because the hadronisation corrections are 
%especially at low \xbj and \Qsq large and model dependent.

The observation of an increased rate of forward jets was 
proposed~\cite{Mueller} as a signature of BFKL--like parton evolution. 
This method asks for jets with a high longitudinal momentum fraction
$x_{jet}=p_{z}/E_{proton}$ %of the proton
(where the parton density functions are well known), for low \xbj (where
the phase space for parton radiation along the gluon ladder is large) 
and for $E^2_{T,jet}$ close to \Qsq (which suppresses the
DGLAP parton evolution).
%$\to$ large $x_{jet}$ pdf well known
%$\to$ low $x$: large phase space for parton radiion long ladders
%$\to$ $\Qsq ~ \et$ suppress DGLAP evolution
%Data: increasing cross-section towards low \xbj  
The ZEUS collaboration~\cite{Riveline} finalised their results on forward jet 
cross-sections. They found an excess at low $x_{Bj}$ compared to 
what DGLAP based LO QCD models like HERWIG~\cite{HERWIG} or LEPTO~\cite{LEPTO} 
predict (see Fig.~\ref{fig:Riveline_1}). 
The ARIADNE~\cite{ARIADNE} model,
which incorporates one of the main features of the BFKL dynamics, namely the 
absence of the $k_T$--ordering in the parton cascade, describes the data
reasonably well. Whether this is purely an effect from $k_T$ non-ordering or 
due to other effects~\cite{rathsman} remains to be investigated in detail.
The LO QCD model RAPGAP~\cite{RAPGAP} adding resolved photon contribution 
to shift the hard subprocess towards the central rapidity region, 
is also able to 
describe the forward jet cross-section~\cite{jungresolved}. 
%Other DGLAP based Models, LEPTO and HERWIG, show a significantly too low
%cross section and flatten out towards lower \xbjx, while data and ARIADNE 
%tend to keep on rising.
%%%%%%%%%%%%%%%%%%%%%%%%%%%%%%%%%%%%%%%%%%
 \begin{wrapfigure}[14]{l}{6.0cm}
\vspace{-0.9cm}
\epsfig{figure=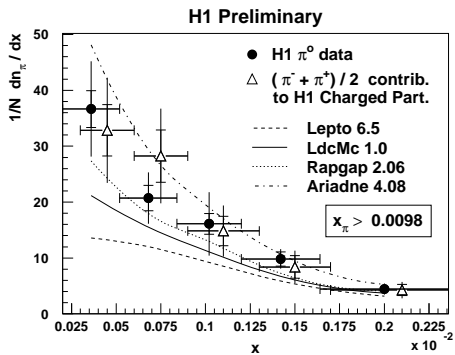,width=7.0cm}
\vspace{-1.1cm}
\caption{Forward pion rate as function of \xbjx. %~\cite{Wengler}.
        QCD models are overlaied for comparison.}
\label{fig:Wengler_1}
\end{wrapfigure}
%%%%%%%%%%%%%%%%%%%%%%%%%%%%%%%%%%%%%%%%%%%%%

An analytical BFKL calculation~\cite{bartels} predicts the rise 
of the cross-section
towards small \xbjx, but is higher than the data.
These calculations are carried out in LO and
the absolute cross-section normalisation is not well known. 
Moreover, hadronisation corrections can be large, especially at
low \xbj and are very model dependent.

%BFKL--based calculations~\cite{lowx:fwdcal} predict too high cross 
%sections. However, it is known that they use approximations, which result
%in an upper limit of the cross section. In addition the normalisation of 
%the absolute cross section in some of the calculations is not well defined.
A general disadvantage of these calculations is the lack of a jet algorithm,
which has to be applied in the experimental analyses to find the jets.
Now new calculations~\cite{Salam,Stirling} 
%--for the first time on forward
%jet production at low \xbjx-- 
have shown that the predicted cross-section
can easily change by a factor two when kinematic constraints, like 
energy-momentum conservation, are applied.
%
%Further new theory calculations -for the first time on forward jet 
%production at low $x_{Bj}$ including BFKL effects- are presented by 
%\cite{Stirling}. They show the importance of kinematic cuts on the selected 
%jets and phase space. 
%
%However, RAPGAP is not using the BFKL equations 
%in the parton shower simulation.
%Further studies on forward pion production by H1 \cite{wengler} follow 
%the expectations from BFKL predictions and show an excess of particle 
%production at low \xbj (see Fig.~\ref{fig:Riveline_1}).
%$\to$ difficult measurement
%$\to$ possibility to calculate fragmentation function
%$\to$ LDC ? figures ?

%%%%%%%%%%%%%%%%%%%%%%%%%%%%%%%%%%%%%%%%%%
\begin{wrapfigure}[18]{r}{6.0cm}
\vspace*{-0.7cm}
\epsfig{figure=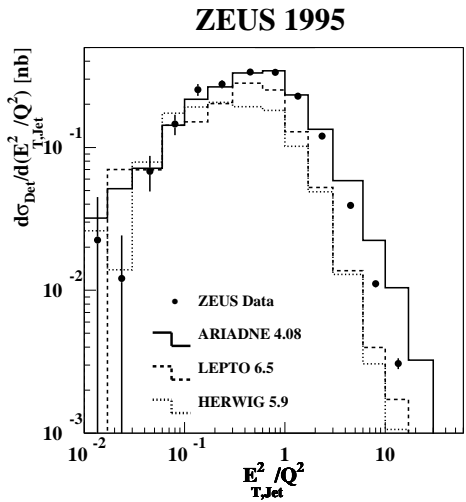,width=7.0cm}  
\vspace*{-0.8cm}
\caption{Forward jet cross section % ~\cite{Riveline} 
         (detector level) 
         as a function of $E^2_{T,jet}/Q^2$.}
\label{fig:Riveline_2} 
\end{wrapfigure}
%%%%%%%%%%%%%%%%%%%%%%%%%%%%%%%%%%%%%%%%%%%%%

A complementary approach to gain insights in the parton evolution
is to study the production of hard particles 
in the central rapidity region\cite{kuhlen}. 
Charged or neutral particles at high \et are well correlated to 
parton activity. Their production rate can be directly
predicted by analytic calculations folded with proper fragmentation
functions~\cite{lang}. The rate of $\pi^0$ and $\pi^\pm$ also 
exhibits a rise towards small \xbj 
(see Fig.~\ref{fig:Wengler_1})\cite{wengler}. 

An important step to improve our understanding of low--$x$ physics is the
Linked Dipole Chain (LDC) model~\cite{LDC}.
A Monte Carlo implementation of this model~\cite{Lonnblad}
was presented for the
first time at this workshop. It allows to simulate the full hadronisation 
on top of a CCFM--based parton evolution. 
Here the trick, which makes the CCFM--type parton evolution 
in a QCD Monte Carlo applicable, is to move as many as possible initial 
state branchings to the final state of the shower.  
However, the prediction of the first released version is below
the data. 
Another approach, even closer to the original CCFM equations, is in 
progress~\cite{Salam}. 
%and will be an important
%consistency check of the hadronisation in the Monte Carlo simulations 
%including explicitly the BFKL/CCFM--based parton shower equations (??).
%
Both approaches could eventually result in a Monte Carlo simulation, 
which sucessfully describes the full 
phase space from photoproduction up to very high \Qsqx. 
How well
the available QCD models behave in such an extended phase space 
can bee seen~\cite{Riveline} in Fig.~\ref{fig:Riveline_2}. 
Here the DIS regime with $E^2_{T,jet} \ll Q^2$ is followed by the
BFKL--like regime, where $E^2_{T,jet} \approx Q^2$, and finally ends in the
region where the hard scale is set by the jet  $E^2_{T,jet} \gg Q^2$. 
While in the DIS regime
all QCD models describe reasonably well the data, in the BFKL regime only
ARIADNE comes close to the data.
In the photoproduction regime all models fail.
In conclusion,
this field needs much more work, before 
firm conclusions can be drawn.
The driving mechanism behind the parton dynamics especially at low \xbj
is not yet understood.

%% file: tevjet.tex
The TEVATRON collider offers the opportunity to investigate the
properties of hard interactions in \pp collisions at distance
scales of approximately $10^{-17}$~cm. The D\O\ and CDF collaborations
presented analyses of the inclusive jet cross-section, the triple
differential dijet cross-section, the dijet mass spectrum, and the
dijet angular distribution~\cite{Bertram} which were all compared
to $\cal{O}($$\alpha_{\rm s}^{3})$ QCD predictions.

As previously reported, CDF observed an excess of high \et
jets with respect to a specific QCD prediction in the inclusive jet
cross-section. Both D\O\ and CDF have released new, preliminary results with
the 1994--95 data sets. The CDF result for jets 
within $0.1 < |\eta| < 0.7$ is in good agreement with the
previous measurement~\cite{cdf_1a_inc} and shows when compared
to EKS NLO calculation~\cite{eks}
an excess of events at high $E_{T}$. 
%
%The updated measurement reduced the systematic
%uncertainties and contains a full calculation of the correlations in the
%errors. 
The JETRAD NLO QCD prediction\cite{jetrad} is, however, in good agreement 
with the  D\O\ data for the rapidity ranges 
$0.1 < \eta < 0.7$ and  $ |\eta| < 0.5$ 
at all values of \etx. In the range  $ |\eta| < 0.5$  
the $\chi^{2}$ for the data-theory
comparison is $23.1$ for $24$ degrees of freedom.
The excess of the high \et jets in the inclusive jet cross-section
found by CDF was often interpreted as a possible sign for a
substructure. Compositeness scales up to $2$~\TeV are, however,
ruled out by an analysis of  dijet angular distribution.

\begin{figure}[htbp]
\vspace{-3.mm}
\begin{minipage}[t]{2.2in}
{\centerline{\psfig{figure=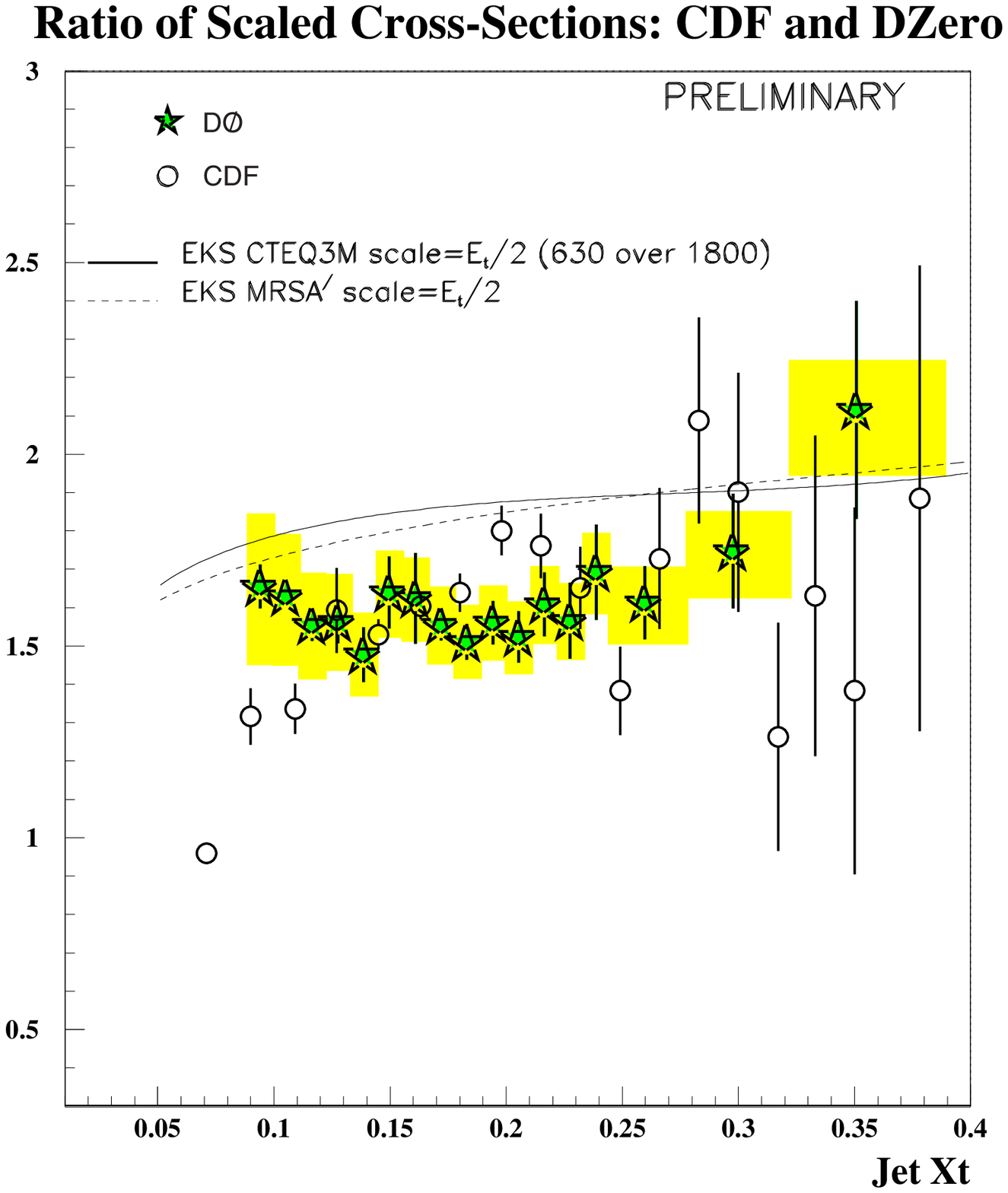,width=2.1in}}}
\vspace{-2mm}
\caption{The ratio of the inclusive jet cross sections at 
\mbox{$\sqrt{s} =$ 1800 and 630 GeV} as a function of $x_{T}$. 
The shaded regions are the systematic uncertainties of the D\O\
result.}
\label{tevatron_fig_3}
\end{minipage}
\hspace{2mm}
\begin{minipage}[t]{2.2in}
{\centerline{\hspace{1mm}\psfig{figure=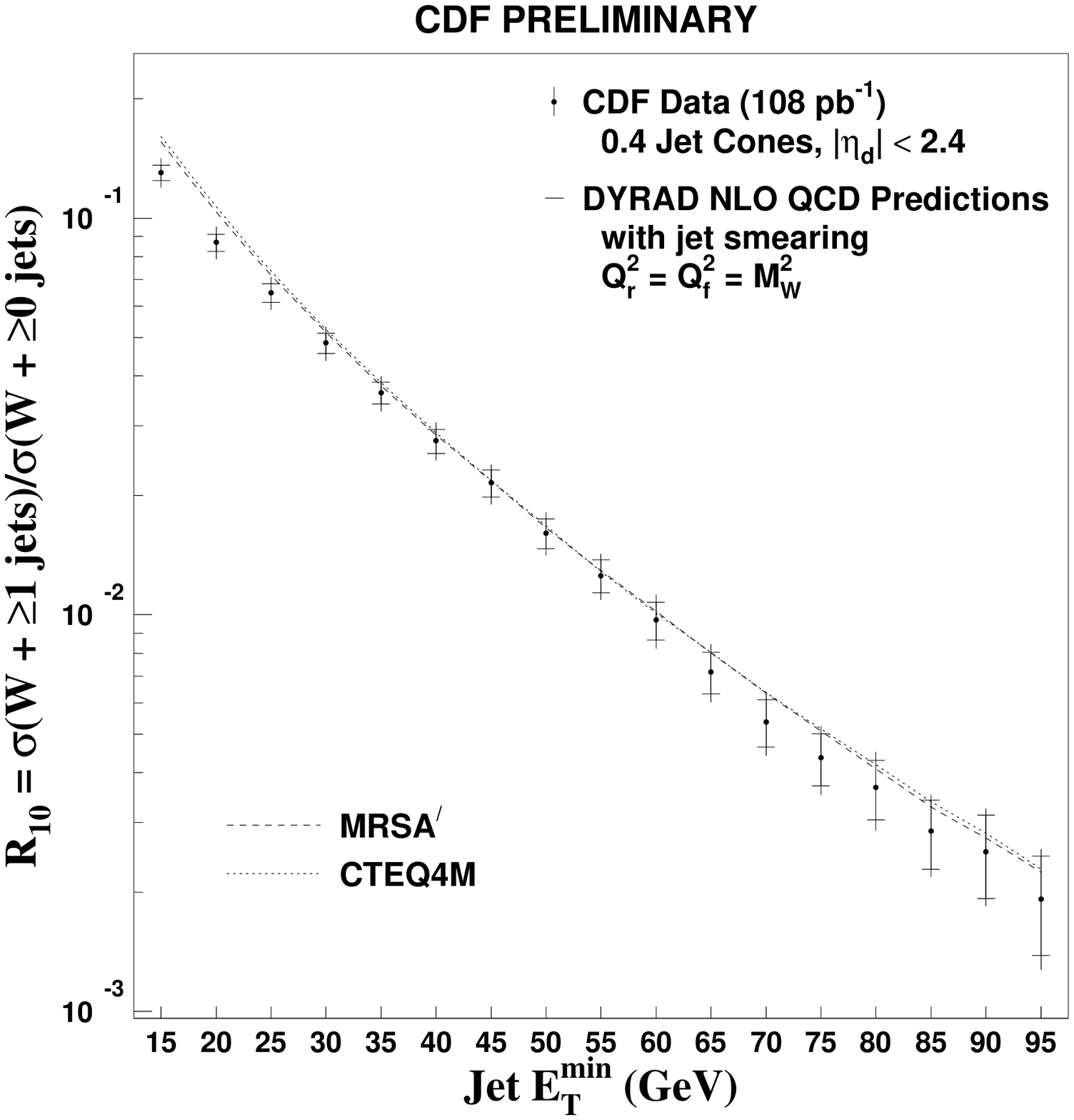,width=2.5in}}}
\vspace{-5mm}
\caption{CDF's measurement of 
$R_{10} \equiv \sigma(W+\ge1\;{\rm Jet})/\sigma(W)$ compared to the
theoretical prediction.}
\label{tevatron_fig_1}
\end{minipage}
\end{figure}

Both experiments also measured the ratio of the inclusive jet cross-section
at \mbox{$\sqrt{s} =$ 1800 and 630 GeV} as a function of \mbox{$x_{T}
= 2 E_{T} / \sqrt{s}$} (as depicted in Fig.~\ref{tevatron_fig_3}). The
data are approximately 15$\%$ below the NLO QCD predictions.

CDF presented a measurement of the dijet triple differential cross
section at \mbox{$\sqrt{s} =$ 1800 \GeVx.} The data are compared to
the JETRAD prediction and show a slight excess of events with
high \et jets. In addition measurements  by both CDF and D\O\ of 
the dijet mass spectrum are
in good agreement with each other and no significant deviation from theory
is seen.

%While CDF sees an excess of jets at high $E_T$ in the inclusive jet
%cross section the dijet angular distribution rules out compositeness
%up to a scale of 2~TeV.
%%%%%%%%%%%%%%%%%%%%%%%%%%%%%%%%%%%%%%%%%%%%%
%%%%%%%%%%%%%%%%%%%%%%%%%%%

%% file: bosjet.tex
 The hadronic interactions of the $W^\pm$ and $Z^0$ bosons provide a clean
 probe of pQCD predictions. Several analyses by D\O\
 and CDF~\cite{Bertram} profit from
 the large sample of $p\bar{p}$ collisions taken during the 1994-96
 collider runs.

 At the DIS97 conference D\O\ presented results~\cite{wjets_dis97}
 which showed an excess of jet production in the exclusive ratio of
 $W^\pm + 1$~jet to $W^\pm + 0$~jet production 
($R_{10}$) when compared to the NLO
 QCD prediction~\cite{dyrad}. The CDF collaboration reported on the
 production cross-section for $W^\pm + \ge n$~jets. The measured 
 cross-sections are a factor of $1.7$ larger than the LO QCD
 predictions~\cite{vecbos}, but are in excellent agreement
 when compared to DYRAD. This is illustrated in Fig.~\ref{tevatron_fig_1},
 where $R_{10}$ as a function of the jet $E_T$ is shown.
 The discrepancy between the D\O\ and CDF results is currently
 unexplained. The most obvious difference between the two
 measurements is that D\O\ uses a cone radius $R=0.7$
 while CDF uses $R = 0.4$. We look forward to the
 updated D\O\ results.

 From the inclusive production
 cross-section for both the $W^\pm$ and $Z^0$ bosons~\cite{Bertram}
  D\O\ extracted the total width of the $W^\pm$:
 \mbox{$\Gamma = 2.126 \pm 0.092 \GeVx$.} 
 A direct measurement of $\Gamma$ by CDF using the tail of the transverse
 mass distribution leads to:
 \mbox{$\Gamma =2.19^{+0.17}_{-0.16}\rm(stat)\pm0.09(syst)\;\rm
 GeV$.}
% Both of these measurements are consistent with Standard Model expectations.
 Both of these measurements give a consistent picture with other electroweak
 measurements within the Standard Model.

 A measurement of the $W^\pm$ and $Z^0$ transverse momentum distributions
 was made by D\O .  In particular the $p_T^Z$ distribution is able to
 distinguish between two available models~\cite{ly,ak} for the
 non--perturbative contributions.

 The NuTeV collaboration presented~\cite{Yu} a new indirect
 measurement of the $W^\pm$ mass: \mbox{$M_{W^\pm} = 80.54 \pm 0.11$
 \GeVx}. D\O\ and CDF presented direct measurements of the $W^\pm$
 mass of: \mbox{$M_{W^\pm} = 80.43 \pm 0.11$ \GeVx} (D\O ), and \\
 \mbox{$M_W^\pm = 80.38 \pm 0.12$ \GeVx} (CDF).

%% file: instanton.tex
In QCD anomalous non-perturbative processes are expected which
violate classical laws like the conservation of chirality.
Instantons~\cite{belavin}, non-perturbative fluctuations of the gluon field
inducing hard processes, represent tunneling
transitions between topologically inequivalent vacua.

In DIS instanton induced processes are dominantly produced in
a quark gluon fusion process~\cite{braun,moch97}. 
The virtuality $Q'^2$ of the quark $q'$,
originating from a photon splitting into a $q\bar{q}$ pair in
the instanton background, provides  
a generic hard scale naturally limiting
the instanton size $\rho$ and makes a quantitative prediction of the
cross-section possible~\cite{moch97}. 
In previous DIS workshops progress on the theoretical understanding
and on possible experimental search strategies~\cite{inststrat} 
has been reported.
First experimental exclusion limits have been derived~\cite{carliins}. 
The maximally allowed fraction of instanton induced events is
 $\cal{O}$ $(1 \%)$.

Recently, further progress on reducing the remaining systematic
uncertainties on the cross-section predictions has been 
achieved~\cite{schrempp}. 
A new $2$ loop renormalisation group invariant calculation of the
instanton density significantly reduces the residual renormalisation
scale dependence of the instanton subprocess cross-section.
Moreover, recent lattice calculations allowed to constrain 
the region of validity of the approximations made 
when deriving the cross-section expected at HERA.
The fiducial kinematic region of $Q'$ and $x'=Q'^2/2 p q'$, 
the Bjorken scaling variable associated with instanton subprocess,
is found to be $Q' \gtrsim 8$\GeV and $x' \gtrsim 0.35$.
For this region and for $\xbjx > 10^{-3}$
and $0.1 < y < 0.9$ the 
instanton production cross-section is found to be 
$\sigma=126$\pbx.
Such a prediction has, however, not yet reached quantitatively 
the predictive power of pQCD.
Given the large data samples corresponding to about $40\pbinv$
per experiment already collected at HERA,
this sizeable cross-section is large enough to pursue dedicated searches
for instanton induced processes. 
The large cross-section of standard DIS processes
necessitates, however, an excellent understanding of perturbative
pQCD models in the tails of distributions. 
The construction of observables based on the characteristic
features of the hadronic final state of instanton induced events,
which allow for a powerful separation from standard DIS processes,
will be the key issue for an experimental discovery.

%% file: evshapes.tex
Measurements of event shape variables
%like thrust, jet mass and jet broadening 
provide information
about perturbative and non-perturbative aspects of QCD.
They allow to fit analytical expressions to data
without referring to a fragmentation model by
exploiting their characteristic power behaviour. % $\cal{O}$ $(1/Q)$.
Event shapes have been extensively studied in 
\ee experiments at different center of mass energies~\cite{bethke,bobbink}.

Results from \ee can be compared to DIS in
the current hemisphere of the Breit frame.
%...(relevant hemisphere enrgies: sqrt(e+e-)/2 and Q/2
%clean separtion of current and target hemishpere
Thanks to the large kinematic range covered at HERA,
the dependence of mean event shape values 
on a hard scale %, like the energy of the scattered
%quark in the Breit frame $Q/2$, 
can be studied in one experiment.
Event shapes are defined\footnote{In the H1 analysis some more,
slightly modified, definitions are studied.} 
by\footnote{The summation of the four-vector 
$p_i = (E_i,\vec{p_i})$ extends 
over all objects $i$ (energy depositions, hadrons or partons)
in the current hemisphere of the Breit frame.}:
\vspace{-0.2cm}
\begin{equation}
   T_Z :=  \frac{\sum_i p_z}
                  {\sum_i |\vec{p}_i|}
   \hspace{0.5cm}
   B_C := \frac{\sum_i |\vec{p}_{Ti}|}
              {2\sum_i |\vec{p}_i|}
   \hspace{0.5cm}
   \rho_C := \frac{M^2}{Q^2} = \frac{(\sum_i p_{i})^2}
                                    {Q^2}
\label{eq:shapes}
\vspace{-0.1cm}
\end{equation}

\begin{figure}[t]
\mbox{\hspace{-0.7cm}
\begin{tabular}{cc}
 \epsfig{figure=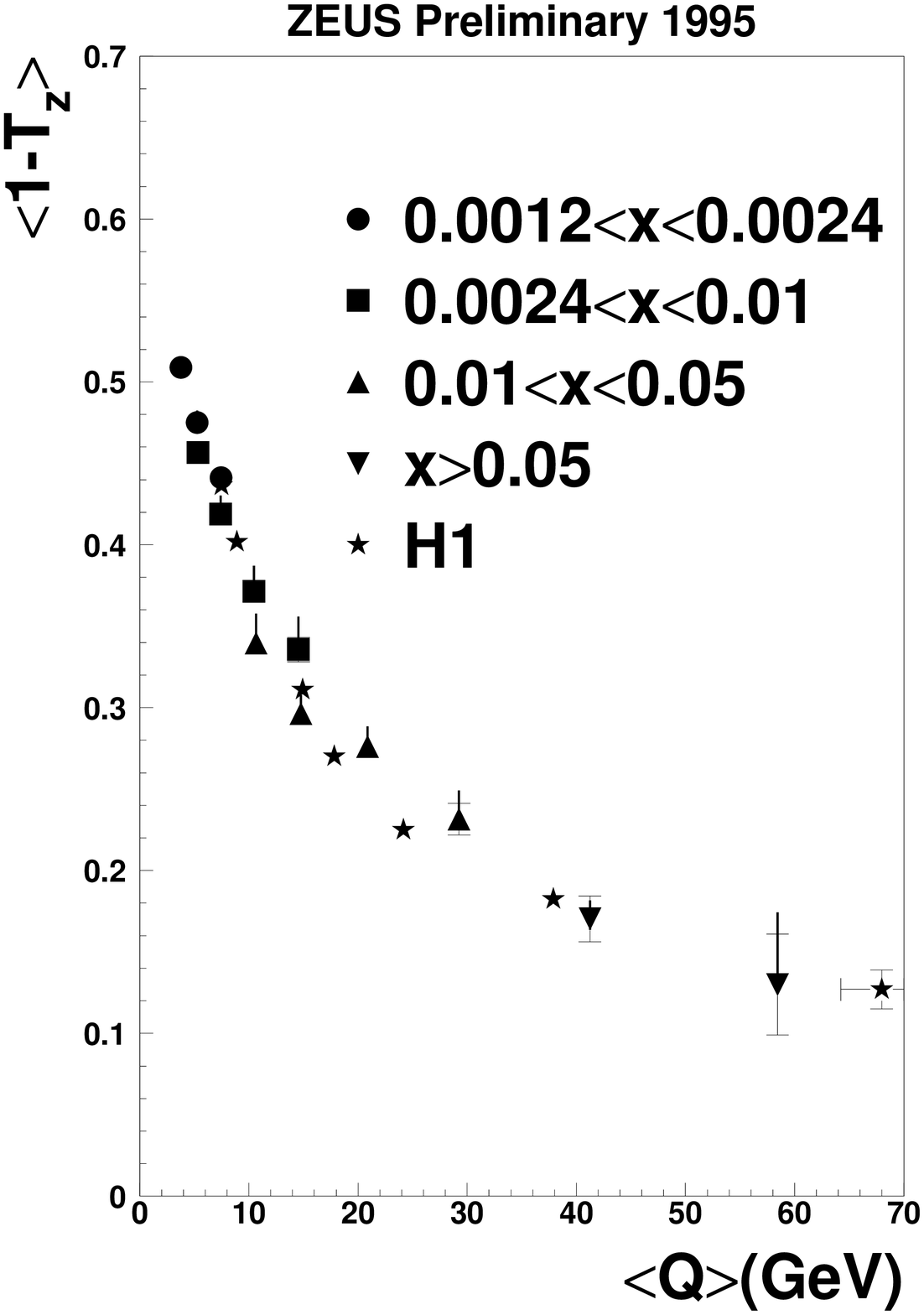,width=5.8cm} &
\mbox{\hspace{-0.7cm}
 \epsfig{figure=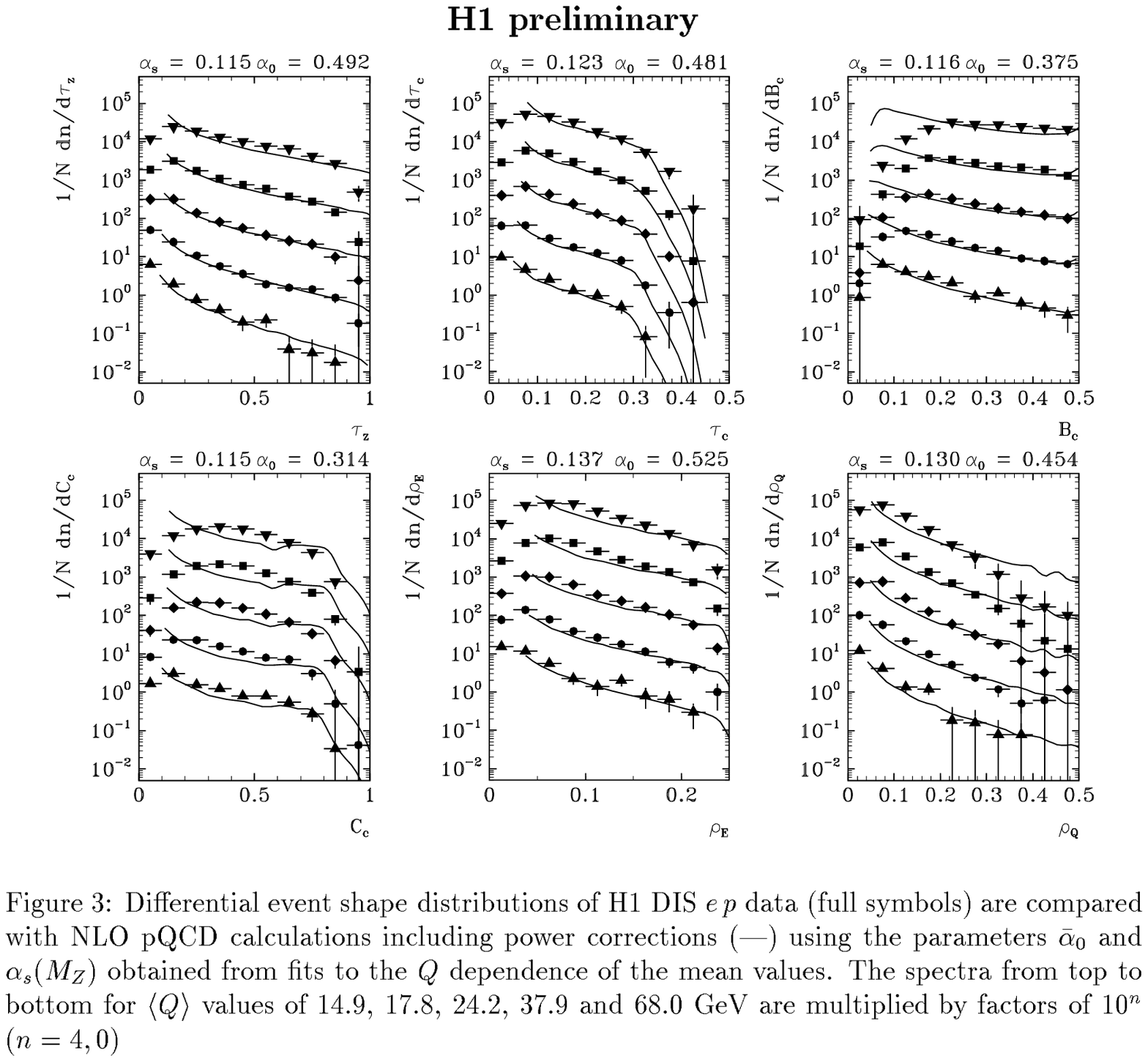,
bbllx=78,bblly=587,bburx=218,bbury=757,clip=,width=6.9cm}
}
\end{tabular}
\begin{picture}(0.,0.)
 {\large
 \setlength{\unitlength}{1.0cm}
 \put(-4.,2.9) {H1 Preliminary}
 }
\end{picture}
}
\vspace{-0.3cm}
\caption{
a) $\av{1-T_Z}$ distribution as a function of $\av{Q}$
   from H1 and ZEUS. %~\cite{Waugh}.
b) Differential $\tau_Z = \av{1 - T_Z}$ distribution for several
   $Q$ bins together with the NLO + power correction theory 
   predictions %~\cite{Martyn} 
   based on the fit parameters obtained from the mean event shapes. 
   The spectra from top to bottom for \av{Q} values
   $14.9, 17.8, 24.2, 37.9$ and $68$\GeV are multiplied by factors
   of $10^n$ for $n=0,...,4$.
}
\label{fig:eventshapes}
\end{figure}

In addition the $C$ parameter 
$C=3 (\lambda_1\lambda_2 + \lambda_2 \lambda_3 + \lambda_3 \lambda_1)$
derived from the eigenvalues $\lambda_i$ of the linearized momentum tensor
is used in the H1 analysis. This analysis~\cite{h1:eventshapes,martyn},
already presented at previous workshops~\cite{wseventshapes},
covers momentum transfers $Q$ from $7$ to $100$ \GeVx.
The ZEUS collaboration has presented first results at this
conference~\cite{waugh} for $3.3 \lesssim Q \lesssim 71.5$ \GeVx.
In the H1 data all particles enter the event shape calculation, 
while ZEUS only selects charged particles.
As an example, the $\av{1-T_Z}$ as function of $Q$
and its differential distribution for different $\av{Q}$ values
are shown in Fig.~\ref{fig:eventshapes}.
Despite the different analysis technique, reasonable agreement
is found.
%%%%%%%%%%%%%%%%%%%%%%%%%%%%%%%%%%%%%%%%%%%%%%%%
%%%%%%%%%%%%%%%%%%%%%%%%%%%%%%%%%%%%%%%%%%%%%
%
A spherical (pencil-like) configuration corresponds to $1-T_Z=1/2$
($0$). The energy flow along the event shape axis
becomes more collimated as $Q$ increases.
Some dependence on \xbj at constant $\av{Q}$ is seen.

The mean event shapes can be expressed by a 
perturbative part calculable in NLO~\cite{mirkes1,disent} 
and a non-perturbative contribution
of the form~\cite{th:powercorr}:
%%%
\vspace{-0.2cm}
\begin{equation}
\av{F}^{pow} = a_F \frac{16}{3 \pi} \frac{\mu_i}{Q}
\ln^p{\frac{Q}{Q_0}} \;
\left [ \bar{\alpha_0}(\mu_i) - \alpha_s(Q) - 
\right ( b \ln{\frac{Q^2}{\mu_i^2}} + k + 2b \left ) \; \alpha_s^2(Q)
\right ] 
\label{eq:fpow}
\vspace{-0.1cm}
\end{equation}
where $b = ( 11 \, C_A - 2 f) /12 \pi$ and $
k = \left [ (67 - 3 \pi^2 ) \, C_A - 10 \, f \right ] / 36 \pi$
and
$\bar{\alpha_0}$ is a free, but 'universal', effective  coupling
parameter below an 'infra-red' matching scale $\mu_i$.
Some of the calculable coefficients $a_F$ and $p$
have recently been reevaluated in the light of 
non-perturbative two-loop corrections
and a common (Milan) factor in eq. \ref{eq:fpow} has been 
introduced~\cite{Salam_2}.
$p$ is only different from $0$ for the $B_C$.
%corrected for detector
%effects without assuming any fragmentation model
%All event shapes variables can be described by such a fit
%(see Fig.\ref{fig:evshape1} for $\av{1-T_C}$).
The power corrections are large at low $Q$, but become less important
with increasing $Q$. 
%For large $Q$ the perturbative
%contribution alone reproduces the shape of the data.
%
$\alpha_s$ and $\bar{\alpha_0}$
can be simultaneously fitted to the data.

QCD fits show~\cite{martyn} that the analytical
form of the power correction is adequate to describe
the data. Only for $B_c$ the theoretically derived additional
factor $\ln{Q/Q_0}$ in eq. \ref{eq:fpow} is not supported by the data.
The 'universality' of $\bar{\alpha_0}$ is
confirmed when using $T_Z$ and $\rho_C$.
The $B_c$ and the $C$ parameter give less consistent
results. Power corrections to differential distributions have not
yet been calculated in DIS. 
A reasonable description of the data can, however, 
be obtained by a simple shift of the distribution like expected
from calculations for \ee.
The resulting fit values are however larger than the ones
obtained from the mean values. 
As illustration, the  $1-T_Z$ is shown in Fig.~\ref{fig:eventshapes}b 
together with the
theory curves predicted from the fit to the mean values.

%% file: fragfun.tex
%%%%%%%%%%%%%%%%%%%%%%%%%%%%%%%%%%%%%%%%%
\begin{wrapfigure}[19]{r}{7.5cm}
\vspace{-0.8cm}
\epsfig{figure=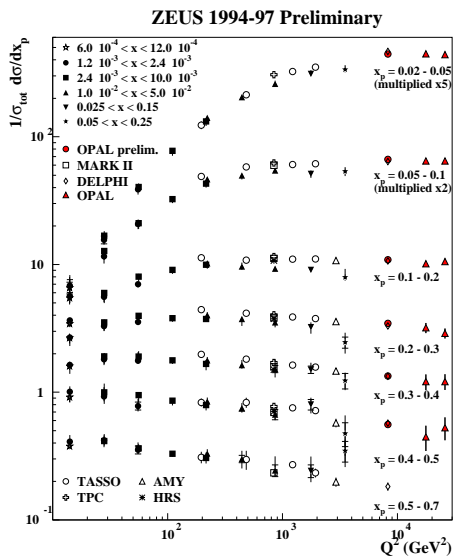,width=7.5cm} 
\vspace{-0.9cm}
\caption{Fragmentation function in $x_p$ bins vs \Qsqx.}
\label{fig:Okrasinski_4}
\end{wrapfigure}
%%%%%%%%%%%%%%%%%%%%%%%%%%%%%%%%%%%%%%%%%%%%%

Fragmentation functions characterize the complete process of hadron 
formation including the QCD parton evolution.
They are most conveniently studied in the current region of the
Breit frame, since there DIS events can be directly compared to 
\ee events in one hemisphere.
As for the proton structure function, scaling violations
are expected.
With increasing \Qsqx, more phase space is available 
and consequently more gluons are emitted. 
%In addition, $\alpha_s$
%gets smaller when probed at a larger scale.
%This results in softer particle spectra. 
To some extent this effect is reduced by the fact that $\alpha_s$ gets
smaller, but the result is a softer particle momentum spectrum.

Therefore more (less) particles are observed at low (large)
scaled momenta $x_p = 2 p / Q$ for higher \Qsqx.
The dependence of $1/\sigma \; d\sigma/dx_p$ on \Qsq for various
$x_p$ bins is shown\cite{Okrasinski} in Fig.~\ref{fig:Okrasinski_4}. 
At intermediate $x_p$ approximative scaling is observed.
At high $x_p$ a violation of this scaling is apparent, 
i.e. less particles with high momentum are found with 
increasing \Qsqx.
The decreasing $1/\sigma \; d\sigma/dx_p$ at low $x_p$ and \Qsq 
is due to mass effects
depopulating the current hemisphere induced by boson-gluon
fusion processes.
The HERA data overlap well within the kinematic range 
of \ee data and good agreement between both types of experiments
is found. 
%First NLO calculations folded with fragmentation functions\cite{fragnlo}
%are available and
%further improvements are in progress to describe the measured data. 
NLO calculations folded with a fragmentation 
function~\cite{fragnlo,binnewies95}
describe the measured distributions well~\cite{Okrasinski} in the
region of high $\Qsqx,x_P$, where the theory is applicable. 
More recent NLO calculations~\cite{fragnlo,binnewies98}
allow to vary the QCD parameter $\Lambda$ and thus will offer the 
possibility to determine $\alpha_s$, when the origin of large fit 
uncertainties to $e^+ e^-$ data is better understood~\cite{Okrasinski}.
Good agreement with the data is found in the high $(x_p,\Qsqx)$
region where the theory is applicable.

Fragmentation functions from individual particle species in jets
produced in $\gamma$p collisions have been measured at HERA\cite{fragjet}.
Their evolution with the scaled energy with respect to the jet energy
is tested.
% and tests the flavor (in-)depencence of the parton branching.
Charged hadrons and kaon production are studied separately. 
The direct component in the photoproduction samples behave very similar to 
the particles in the DIS samples and both are well consistent with results 
from $e^+e^-$ experiments
(see Fig.~\ref{fig:cos_theta} in ref.~\cite{Okrasinski}). 
This supports universality of fragmentation in different reactions, 
if the appropriate observables are studied.

However, the mechanism driving
the transition of partons to hadrons is little understood. 
%The concept of colour confinement is just an excuse (??) 
%for our non-knowledge on how this transition in the 
%non-perturbative regime of QCD takes place. 
In the  perturbative regime some colour related effects like 
soft colour coherence have been successfully calculated,
e.g. the modified leading log approximation
(MLLA)\cite{mlla}. Any effect, with which
one can study low--energy colour interactions helps to understand
the nature of non-perturbative QCD better.
$W$--pair production at LEP~\cite{Orava} is used to investigate
colour recombinations effects. Events with 4-jets are selected as $W^\pm$ pair
candidates. 
%%%%%%%%%%%%%%%%%%%%%%%%%%%%%%%%%%%%%
\begin{wrapfigure}[13]{r}{5.5cm}
\vspace{-1.0cm}
\epsfig{figure=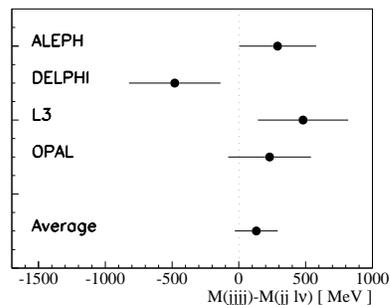,width=6.0cm}
\vspace{-0.9cm}
\caption{Invariant mass difference in $W^\pm$ pair production 
events with $2$ and $4$ jets.}
\label{fig:Orava_1}
\end{wrapfigure}
%%%%%%%%%%%%%%%%%%%%%%%%%%%%%%%%%%%%%%%%

%The invariant mass of two of the jets should reflect the
%$W$--mass. 
If the two proper jets are found, the $W^\pm$ mass can
be reconstructed. Soft colour reconnection could smear out
%If soft gluons are exchanged between partons emerging from different
%$W$'s then 
the width of the $W^\pm$ mass peak. 
%should be smeared out. 
The LEP experiments have searched for this effect, but so far the results 
do not show a statistically significant signal, see Fig.~\ref{fig:Orava_1}.  
Similar conclusions are drawn for other observables
like multiplicity differences or Bose-Einstein correlation effects.
A more sophisticated method~\cite{Orava} tries to reconstruct
the colour flow event by event and might be more sensitive to recombination
effects.